\documentclass[
  11pt,                     			
 a4paper,            			
onecolumn,          			
]{scrartcl}

\usepackage{authblk}
\usepackage{ae}
\usepackage{scrpage2}		
\usepackage{slashed}
\usepackage[utf8]{inputenc}
\usepackage{amsmath}		
\usepackage{amssymb}		
\usepackage[dvips]{graphicx}		
\usepackage{caption}		
\usepackage[active]{srcltx}
\usepackage[dvipdfm]{hyperref}		
\usepackage{exscale}
\usepackage{caption}
\usepackage{subcaption}
\usepackage{float}
\usepackage{array}
\DeclareMathSizes{10.95}{9}{7}{7}
\usepackage{exscale}
\usepackage[dvips]{graphicx}
   \usepackage{amsmath,amsthm}
    \usepackage{amssymb}

\usepackage{ifthen}
\newcommand*{\appendixmore}{%
  \renewcommand*{\othersectionlevelsformat}[1]{%
    \ifthenelse{\equal{##1}{section}}{\appendixname~}{}%
    \csname the##1\endcsname\autodot\enskip}
  \renewcommand*{\sectionmarkformat}{%
    \appendixname~\thesection\autodot\enskip}
}

\newcommand{\qb}{\ensuremath{\overline{Q}}}
\def\A[#1,#2,#3]{\frac{\delta}{\delta A_{#1}^{#2}\left(#3\right)}}
\def\bra#1{\left<#1\right|}
\def\ket#1{\left|#1\right>}
\def\com[#1,#2]{\left[#1,#2\right]}
\def\acom[#1,#2]{\big\{#1,#2\big\}}
\def\bracket[#1,#2]{\left<#1|#2\right>}

\def\vn[#1]{\nabla_#1}

\def\<-#1{\overleftarrow{#1}}
\def\>#1{\overrightarrow{#1}}
\def\nr[#1,#2]{\left(#1\right)^{\left(#2\right)}}

\def\u{\nu}
\def\-1{\nu}
\def\0{\mu}
\def\1{\alpha}
\def\2{\beta}
\def\3{\gamma}
\def\L[#1]{{_#1}}
\def\U[#1]{{^#1}}
\def\ga[#1]{\Gamma \left(#1\right)}
\def\ub{\overline{u}\left(p^{\prime},\sigma^{\prime}\right)}
\def\u{u\left(p,\sigma\right)}

\begin{document}

\title{Heavy quark mass expansion of vector and tensor currents and intrinsic charm in nucleon form factors}
\date{}

\author[1,2]{Maxim V. Polyakov}
\author[1]{Jan Sieverding\thanks{e-mail: Jan.Sieverding@tp2.ruhr-uni-bochum.de}}
\affil[1]{Institut f\"ur Theoretische Physik II, Ruhr-Universit\"at Bochum,
D-44780 Bochum, Germany}
\affil[2]{Petersburg Nuclear Physics Institute, Gatchina, 188300, St.~Petersburg, Russia}


\maketitle

\begin{abstract}
The framework of the expansion by subgraphs is used to compute asymptotic expansions for the vector and the tensor currents in the limit of large quark masses.
We use the results to obtain
an estimate for the influence of heavy quarks  on the nucleon electromagnetic and tensor form factors.  
\end{abstract}

\section{Introduction}
The intrinsic heavy quark content of light hadrons is a fundamental property of QCD \cite{Brodsky:1980pb,Harris:1995jx,Brodsky:2015fna}.
The influence of intrinsic heavy quark content on the structure of the nucleon can be investigated by several methods. Much information about the nucleon structure like the 
charge distribution and the anomalous magnetic moment can be obtained from form factors. These form factors parametrize the expectation values of bilinear fermion operators 
in a single-nucleon state. To describe the influence of intrinsic heavy quarks on the form factors one should therefore consider those bilinear fermion operators 
that contain heavy quark fields, it is natural to employ a large mass expansion for the investigation of such operators.
One way to perform large mass expansions is the heavy quark mass expansion (HQME) which was used in \cite{Franz:2000ee} and \cite{vc2009}  to obtain an expansion of 
the vector and the tensor current of heavy quarks in terms of gluon field operators.
\par
One observation that the HQME-approach is not complete was made in \cite{Ji:2006vx}. This observation concerns the influence of muons on the anomalous magnetic 
moment $\mu_e$ of the electron in QED. If the muon mass is treated as large parameter (compared to the electron mass) this influence can be described by using the HQME to expand the muonic vector current
in terms of photon field operators. The first photon field operators in the expansion that contribute to $\mu_e$ are of order $1/m_\mu^4$ \cite{Kaplan:1988ku}. However, from the evaluation of the corresponding Feynman 
diagrams one can show \cite{Laporta:1991zw} that the contributions of muons to $\mu_e$ start at order $1/m_\mu^2$. Due to the similarity of the calculations, the same problems can be expected 
for the HQME of the vector current of heavy quarks in QCD.
In \cite{Ji:2006vx} it was argued that the problem in QED arises because the $1/m_\mu^4$-suppressed photon operator has to be renormalized by  $1/m_\mu^2$-suppressed electron operators. That
settles the problem in this particular case, but it would be desirable to have a systematic procedure to find those additional operators that complete the operator product expansions of the HQME. 
\par 
To give such a systematic procedure, we use the language of Feynman diagrams instead of that of path integrals that was used to derive the HQME results. 
In this language, rigorous results about the influence of virtual heavy particles are known for a long time.
In Ref.~ \cite{Appelquist:1974tg} the decoupling theorem was proved: one can find renormalization schemes, in which, to leading order in the mass of the heavy particles, 
the effects of the heavy degrees of freedom can be neglected altogether.
Afterwards much work was devoted to the problem of how to obtain a systematic expansion that allows to take into account the effects of heavy particles both on the diagrammatic level and on 
the level of the Green functions  (see for example Ref.~ \cite{Kazama:1979xc}, Ref.~ \cite{Witten:1975bh}). The first treatments were done within a momentum
subtraction scheme because here the close connection between asymptotic expansions and renormalization was most apparent. 
However, due to the fact that dimensional regularization is nowadays the 
preferred method for regularization both because of its technical simplicity and its preservation of gauge invariance, it was desirable to derive similar results within the minimal subtraction scheme.
In the MS-scheme decoupling is not straightforward: not all effects of heavy particles are suppressed by powers of their masses. 
Instead, one has to absorb effects of heavy particles in the physical coupling constants of the theory \cite{Bernreuther:1981sg}. Subsequently a systematic procedure for the asymptotic expansions
within the MS-scheme was developed (see for example \cite{Smirnov:1990rz}). Finally the technique of `expansion by subgraphs' \cite{Smirnov:2002pj}  was developed, for which it was proved 
that one obtains an expansion that is free of infrared divergencies and that has the correct $1/M$ behaviour. This technique works on the diagramatic level in the first place. However, 
with the help of the counterterm technique these results could be immediately generalized to an operator product expansion.
\par
Here we will use this technique to analyze the validity of the HQME-approach. This is done for the especially important examples of the vector current and the tensor current. 
The Section explains the basic idea of  asymptotic expansions.
In the following section the expansion is carried out for the vector current and the tensor current. 
Finally, we use the expansion of the vector current to estimate the influence of charm quarks on the e.m. and tensor form factors of the nucleon.
The reader interested in the final result can directly go to Section~\ref{sec:result}. Applications for the intrinsic charm in nucleon form factors are discussed in Sections~\ref{sec:appl1} and
\ref{sec:appl2}.

\section{Asymptotic expansions in heavy quark masses}
\label{sec:asexp}
Perhaps the simplest way to illustrate the general idea is to cite the power counting theorem for the dependence of a Feynman diagram on the mass of a heavy particle 
from Ref.~\cite{Caswell:1981xt}:
\begin{quote}
Let $\Gamma$ be a diagram with $l$ loops. The asymptotically irreducible diagrams (AI-diagrams) of $\Gamma$ are those connected subdiagrams that contain heavy
lines and cannot be made disconnected by cutting a single light line. Then the asymptotic 
behaviour of $\Gamma$ is bounded by
\begin{equation*}
  m^{\sum_{\gamma \in S} deg\left(\gamma\right)}\;\log\left(m\right)^l
\end{equation*}
where $S$ is that set of disconnected AI graphs (spinney) containing all heavy lines which has the highest degree of divergence.
\end{quote}
From this it is immediately clear that by introducing an operator that
lowers the degree of divergence of the AI-diagrams sufficiently (like the R-operation in momentum subtraction schemes lowers the degree of divergence to make 1PI graphs convergent) one can generate a remainder that is
suppressed by an arbitrarily chosen power of the heavy mass. The difference between this remainder and the original (renormalized) diagram will then constitute the correct asymptotic expansion 
for the diagram.\par
This procedure is systematically described  in \cite{Smirnov:1990rz}. We only give the main result.\\ We first have introduce some notation.
We denote the large mass in which the expansion is done by $M$, the external momenta are collectively denoted by $\{p\}$. Let $\Gamma$ be an arbitrary graph 
and let the diagram $I_\Gamma(\{p\},m,M)$ be the corresponding analytic expression. The degree of divergence of the 
diagram $I_\gamma$ is $\omega_\gamma$. 
Let $S_{AI}(\Gamma)$ be the set of all AI-spinneys (set of mutually disjoint AI subgraphs) and $I_{\Gamma/S}$ the diagram $I_\Gamma$ with all elements of the spinney $S$ shrunk to a point. 
$I_{\Gamma/S}\circ \prod_{\gamma \in S}V_\gamma$ is the diagram in which the vertex $V_\gamma$ replaces the subgraph $\gamma$ for each $\gamma\in S$.  
$R^{un}I_{\Gamma/S}$ is the R-operation that only acts on those parts of the diagram that do not contain vertices to which the graphs of $S$ were shrunk. Finally,
$M^{a_\gamma}_\gamma$ performs a Taylor expansion of the diagram $I_{\gamma}(\{p\},m,M)$ in the momenta $\{p\}$ external to the graph $\gamma$ and the light masses $m$ up to the order $a_\gamma$. \\
In \cite{Smirnov:1990rz} it was shown that for $a_\gamma=a+\omega_\gamma$ one has, up to $\mathcal{O}\left(\frac{1}{M^{a+1}}\right)$-terms,
\begin{align}
\label{eq:AE_general}
R I_{\Gamma}(\{p\},m,M)\sim \sum_{S \in S_{AI}(\Gamma), S \neq 0}R^{un}I_{\Gamma/S}(\{p\},m)\circ \prod_{\gamma \in S} M^{a_\gamma}_\gamma \;R I_{\gamma}(\{p\},\{q\},m,M)\textnormal{.}
\end{align}
It was also proven that this result is free of artificial UV- and IR-divergences, that is, the divergences of this expression are the same as those for the original diagram. To summarize:
To perform an asymptotic expansion in large masses on has to find all AI-spinneys of a diagram $\Gamma$. In each subgraph $\gamma$ of this spinney, one performs a Taylor expansion in external momenta and light masses.
The result is a new vertex factor $V_\gamma$ that is reinserted in the original diagram to replace the subdiagram $\gamma$. Then the unaffected parts of $\Gamma$ are renormalized as usual.\par
The transition from the diagrammatic to the operator level is now fairly obvious: If one computes the matrix element of an operator containing heavy particles one will obtain Feynman diagrams 
containing AI-subgraphs. The expansion of these diagrams in terms of the heavy mass amounts to replacing the AI subgraphs by their Taylor expansions in light masses and 
external momenta. The terms of this expansion serve as new effective vertices in the complete diagram. The complete dependence on the heavy mass is in these effective vertices.
Thus, if we reexpress the effective vertices as vertices due to  local operators in the light degrees of freedom, all
matrix elements  of the heavy operator can be described by matrix elements of light operators.\\
Now, if one wants to expand the heavy operator $\mathcal{O}$ of dimension $dim(\mathcal{O})$ to the order $a$, on the diagrammatic level one has to expand each AI-subgraph to the 
order $a_\gamma=\omega_\gamma+a$. Since in theories with dimensionless coupling constant $\omega_\gamma=dim(\mathcal{O})-E_B-\frac{3}{2}E_F$ ($E_B$, $E_F$ denote the number of external boson- resp.
fermion-propagators), it is enough to consider diagrams with $0\leq a+ dim(\mathcal{O})-E_B-\frac{3}{2}E_F$. For example, at the order $1/M^2$ for  $dim(\mathcal{O})=3$
only diagrams with  $E_B \leq 5$ and $E_F \leq 2$ (remember that external fermion lines have to come in pairs) have to be taken into account. In gauge theories, Ward-identities will lower 
the actual degree of divergence for diagrams with external vector bosons. This reduces the number of diagrams that have to be considered.\\
In this paper the vector current and the tensor current are considered. For these operators an OPE is done to the order $1/M^2$ for the vector current
and to the order $1/M$ for the tensor current and up to the order $\alpha^3$ in the strong coupling constant. The reason for the order $\alpha^3$ is the fact that this is the lowest order 
with purely fermionic operators.

\section{The AI-diagrams for fermion-bilinears}\label{sec:AI-diagrams}
We first give a general constraint for the structure of the expansion (see the appendix of Ref.~ \cite{Kazama:1979xc}). 
For the argument that follows we use a regulator that admits the usual definition  of $\gamma_5$.
Since the heavy particles should not occur as external particles, every AI-diagram will contain heavy fermions only inside closed loops. 
If we switch the sign of $M$, we have
\begin{align*}
S_{F}\left(p,-M\right)=\quad & -\gamma^{5}S_{F}\left(p,M\right)\gamma^{5}\textnormal{.}
\intertext{Now, the factors of $\gamma^5$ can be absorbed in the vertex factors via}
-\gamma^{5}\gamma^{\mu}\gamma^{5}=\quad & \gamma^{\mu}\\
-\gamma^{5}\sigma^{\mu\nu}\gamma^{5}=\quad&-\sigma^{\mu\nu}\\
-\left(\gamma^{5}\right)^3=\quad&-\gamma^5 \textnormal{.}
\end{align*}
This cancels all $\gamma^5$ in the loop, leaving the diagrams with vector current insertion as they are and changing the sign of the diagrams with tensor 
current insertion. One can conclude that in the expansion of the vector current only
even powers of $M$ can occur while the expansion of the tensor current contains only odd powers of $M$.\par
We can further restrict the types of relevant diagrams.
Only gluons can couple directly to the heavy quark loop. This means that each AI diagram will have a subdiagram that can be considered as off-shell matrix element of
\begin{align*}
G_{n}^{\mu_1,\dots,\mu_n}\left(x,p_1,\dots,p_n\right):=\quad \bra{0}T\mathcal{O}(x)A^{\mu_1}\left(p_1\right)\dots A^{\mu_n}\left(p_n\right)\ket{0}\textnormal{.}
\end{align*}
Matrix elements with one external gluon are zero by color conservation. Let us consider the case of two external gluons. Under charge conjugation, we have
\begin{align*}
\overline{\Psi}\gamma^\mu\Psi\quad\xrightarrow{C}&\quad -\overline{\Psi}\gamma^\mu\Psi\\
\overline{\Psi}\sigma^{\mu\nu}\Psi\quad\xrightarrow{C}&\quad -\overline{\Psi}\sigma^{\mu\nu}\Psi\\
A_{\mu}\quad\xrightarrow{C}&\quad -A_{\mu}^T\textnormal{.}
\end{align*}
Thus, using global color conservation results in
\begin{align*}
G_{2}^{\mu_1,\mu_2}\left(x,p_1,p_2\right)\quad&=\quad \bra{0}T\mathcal{O}(x)A_a^{\mu_1}\left(p_1\right) A_b^{\mu_2}\left(p_2\right)\ket{0}\\
\quad&=\quad \frac{\delta^{a\,b}}{4}\bra{0}T\mathcal{O}(x)tr_c\left(A^{\mu_1}\left(p_1\right) A^{\mu_2}\left(p_2\right)\right)\ket{0}\\
G_{2}^{\mu_1,\mu_2}\left(x,p_1,p_2\right)\quad&\xrightarrow{C}\quad (-1)\frac{\delta^{a\,b}}{4}
                                                  \bra{0}T\mathcal{O}(x)tr_c\left(\left(A^{\mu_1}\right)^T\left(p_1\right)\left(A^{\mu_2}\right)^T\left(p_2\right)\right)\ket{0}\\
                                     \quad&= -\quad \frac{\delta^{a\,b}}{4}\bra{0}T\mathcal{O}(x)tr_c\left(A^{\mu_1}\left(p_1\right) A^{\mu_2}\left(p_2\right)\right)\ket{0}=0\textnormal{.}
\end{align*}
This shows that for the vector current and the tensor current one needs to consider only diagrams where the fermion loop is connected to the rest of the diagram by at least three gluon propagators.\\
One can also find the only possible color structure for the matrix elements with three external gluons: From global color conservation we see that there 
are only two relevant color structures for the matrix element of the current with three external gluons ($8\otimes8\otimes 8$ contains two different singlets). These can be taken 
as the two independent trace structures. In combination with charge conjugation invariance this leads to:
\begin{align*}
G_{abc}^{\mu_1,\mu_2,\mu_3}\left(x,p_1,p_2,p_3\right)\quad=&\quad \bra{0}T\mathcal{O}(x)A_a^{\mu_1}\left(p_1\right) A_b^{\mu_2}\left(p_2\right)A_c^{\mu_3}\left(p_3\right)\ket{0}\\
                                                 \quad=&\quad -\frac{i}{12}\;f^{abc}\;\bra{0}T\mathcal{O}(x)tr_c \left(A^{\mu_1}\left(p_1\right) \com[A^{\mu_2}\left(p_2\right),A^{\mu_3}\left(p_3\right)]\right)\ket{0}\\
                                                      \phantom{=}&\quad +  \frac{3}{20}\;d^{abc}\;\bra{0}T\mathcal{O}(x)tr_c \left(A^{\mu_1}\left(p_1\right) \acom[A^{\mu_2}\left(p_2\right),A^{\mu_3}\left(p_3\right)]\right)\ket{0}\\
\quad \xrightarrow{C}&\quad \phantom{-}\frac{i}{12}\;f^{abc}\;\bra{0}T\mathcal{O}(x)tr_c \left(A^{\mu_1}\left(p_1\right) \com[A^{\mu_2}\left(p_2\right),A^{\mu_3}\left(p_3\right)]\right)\ket{0}\\
                                                               \phantom{=}&\quad+\frac{3}{20}\;d^{abc}\;\bra{0}T\mathcal{O}(x)tr_c \left(A^{\mu_1}\left(p_1\right) \acom[A^{\mu_2}\left(p_2\right),A^{\mu_3}\left(p_3\right)]\right)\ket{0}\\
\quad\overset{!}{=}&\quad G_{abc}^{\mu_1,\mu_2,\mu_3}\left(x,p_1,p_2,p_3\right)\textnormal{.}
\end{align*}
Thus,
\begin{align*}
\bra{0}T\mathcal{O}(x)tr_c \left(A^{\mu_1}\left(p_1\right) \com[A^{\mu_2}\left(p_2\right),A^{\mu_3}\left(p_3\right)]\right)\ket{0} \quad=\quad 0\textnormal{.}
\end{align*}
For the case of the vector current one can conclude from QED current conservation that there is an additional factor of the momentum entering the diagram at the operator insertion. By dimensional 
analysis, the dimension in the heavy mass is lowered by by $1$.\\
We will show in section \ref{sec:loop_structure} that if three gluons couple directly to the fermion loop each of the external gluons comes with a factor of external
momentum. Taking into account both kinds of momentum factors, we have the power in the heavy mass $\overline{\omega_\gamma}=\omega_{\gamma}-E^h_g-\delta$ where $E^h_g$ is the number of external gluons 
coupled directly to the heavy loop, $\delta=1$ for the vector current and $\delta=0$ for the tensor current \textbf{if} no more than three gluons couple directly to the heavy loop. If there are more than three
gluons coupled to the loop, we have $\overline{\omega_\gamma}=\omega_{\gamma}-\delta$.\par
\begin{figure}
  \centering
\subcaptionbox{1 loop: type I\label{1loop}}
  [.2\linewidth]{\includegraphics[scale=.15]{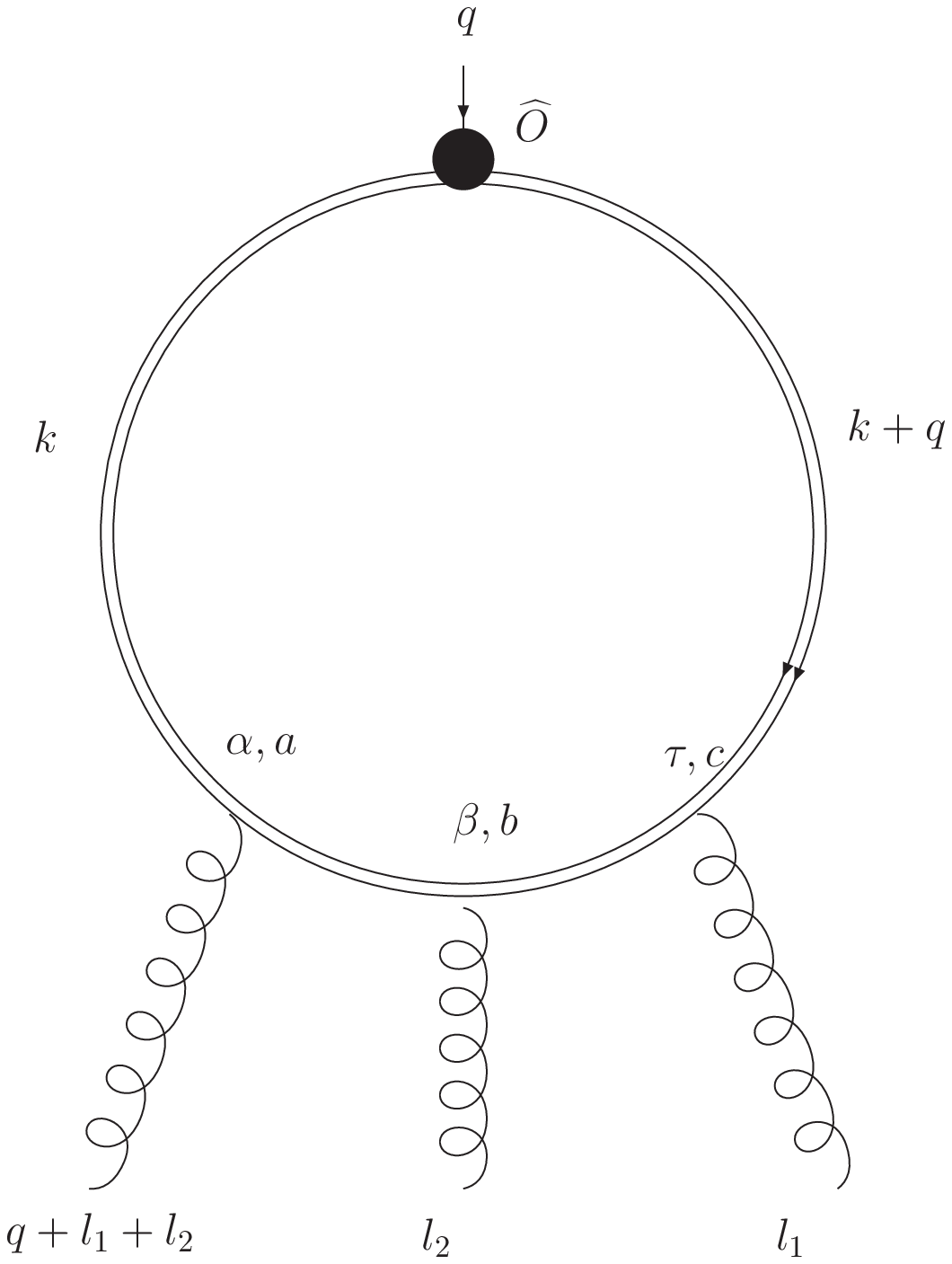}}
\subcaptionbox{1 loop: type II\label{1loop2}}
  [.2\linewidth]{\includegraphics[scale=.15]{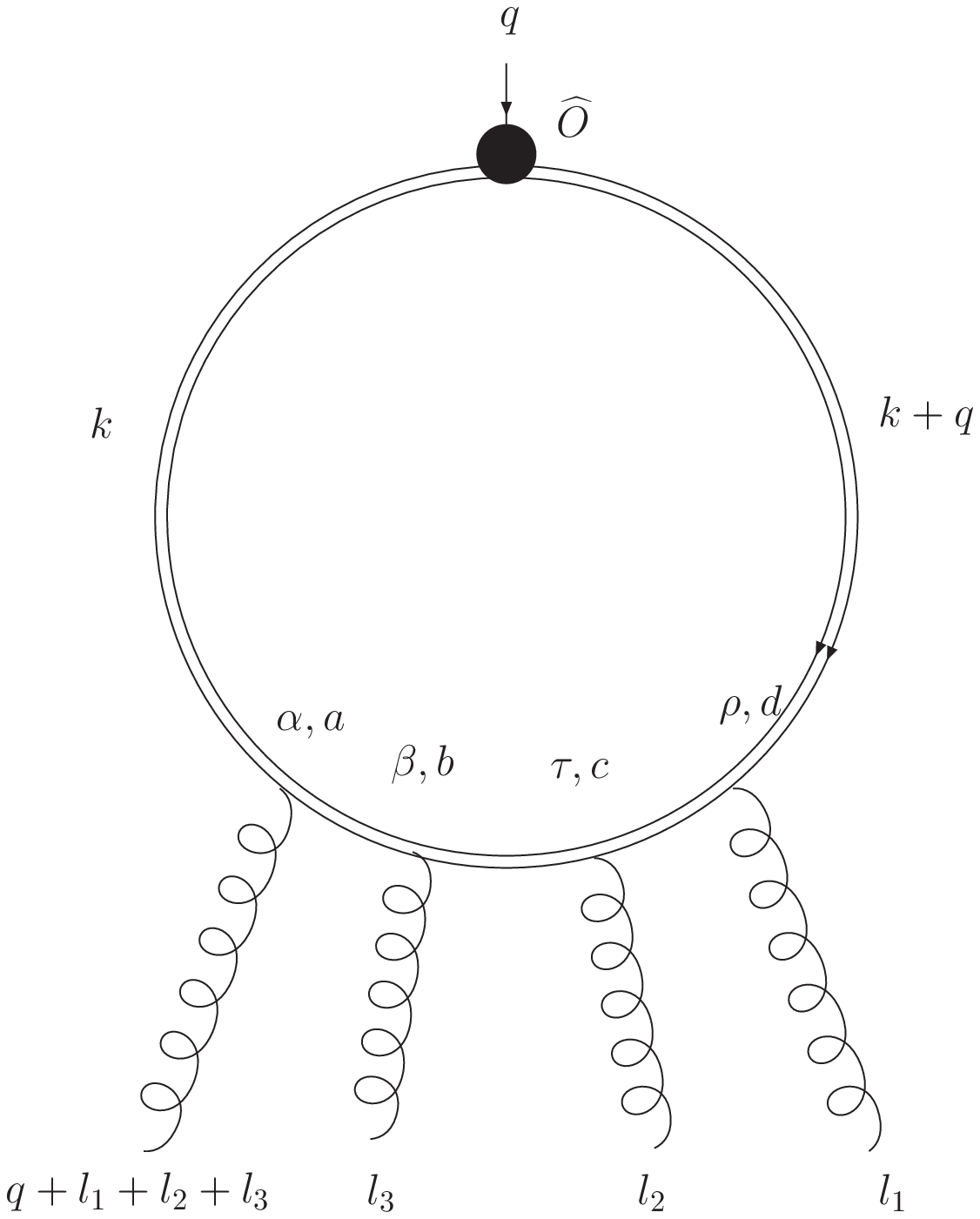}}
\subcaptionbox{2 loop: type I\label{2loop3gluon}}
  [.2\linewidth]{\includegraphics[scale=.15]{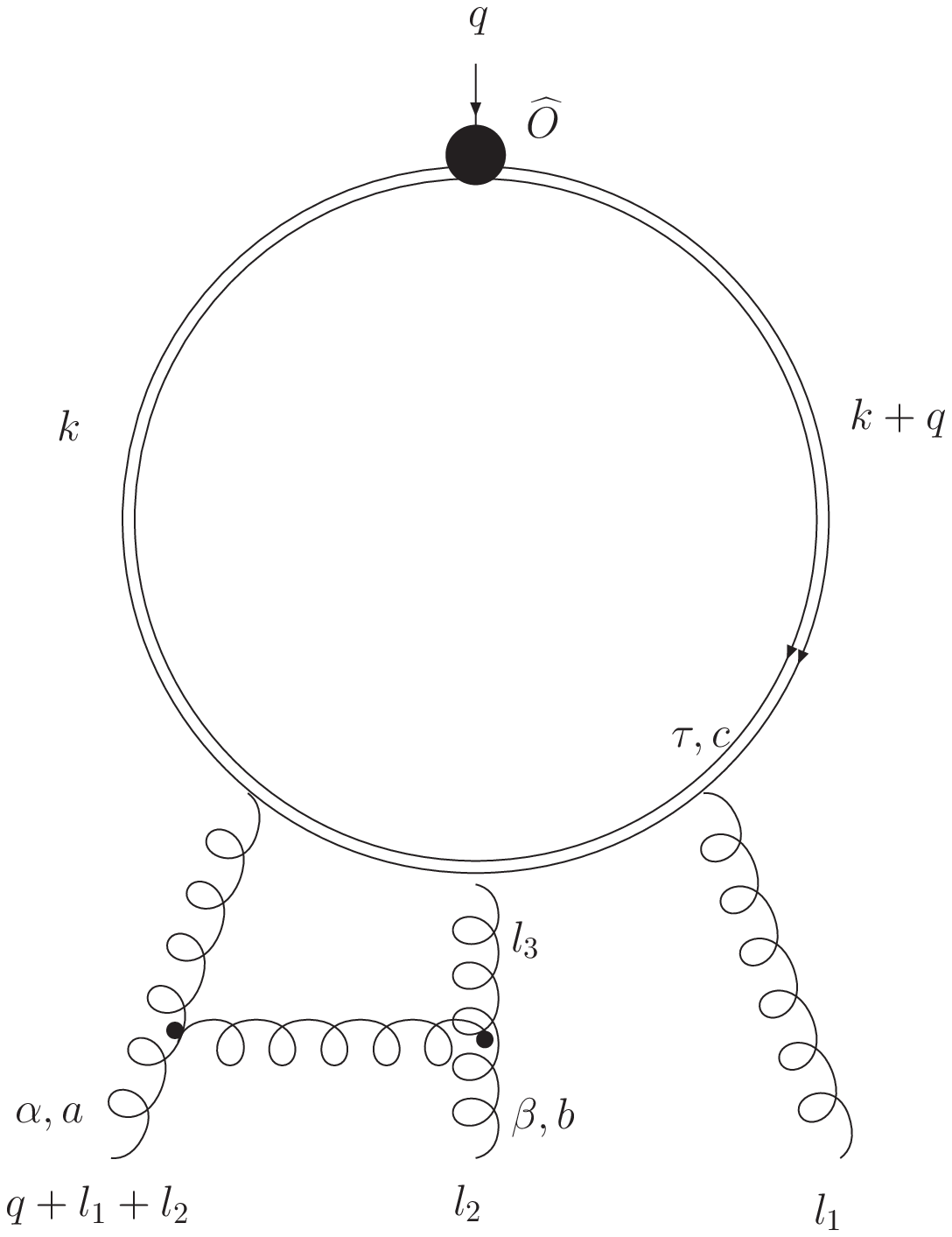}}
\subcaptionbox{2 loop: type II\label{2loop}}
  [.2\linewidth]{\includegraphics[scale=.15]{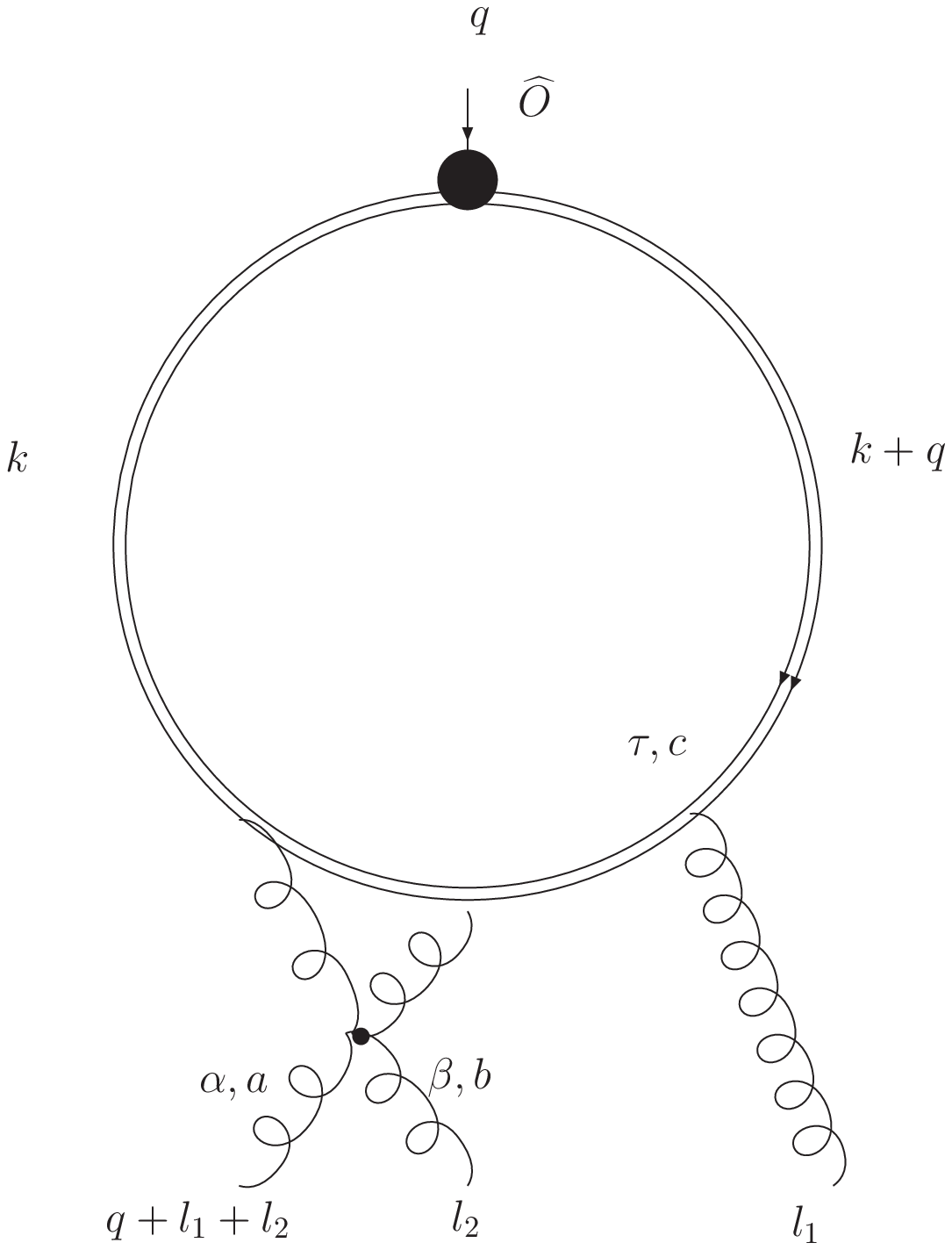}}
\subcaptionbox{2 loop: type III\label{2loop4gluon}}
  [.2\linewidth]{\includegraphics[scale=.15]{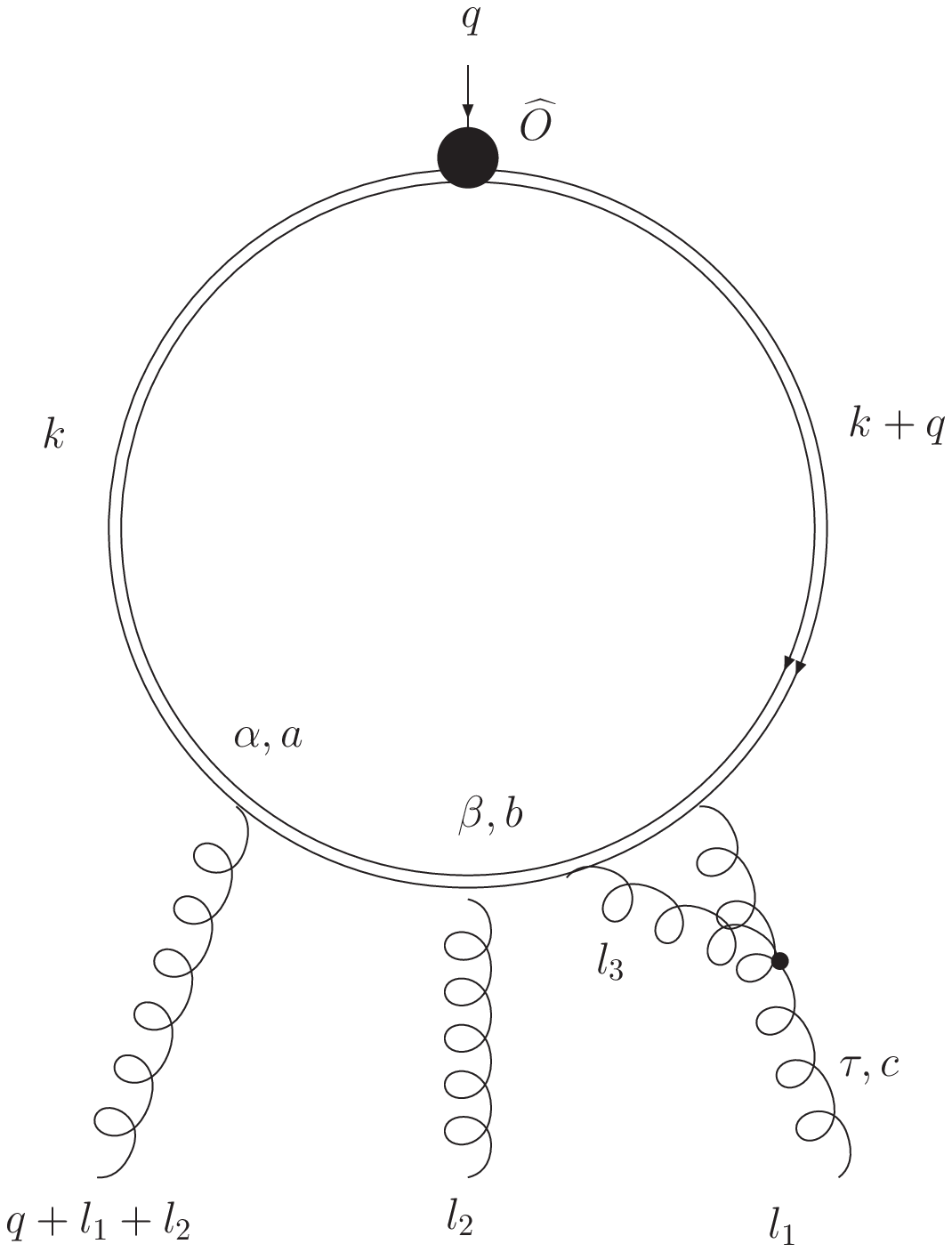}}
\subcaptionbox{2 loop: type IV\label{2loop5gluon}}
  [.2\linewidth]{\includegraphics[scale=.15]{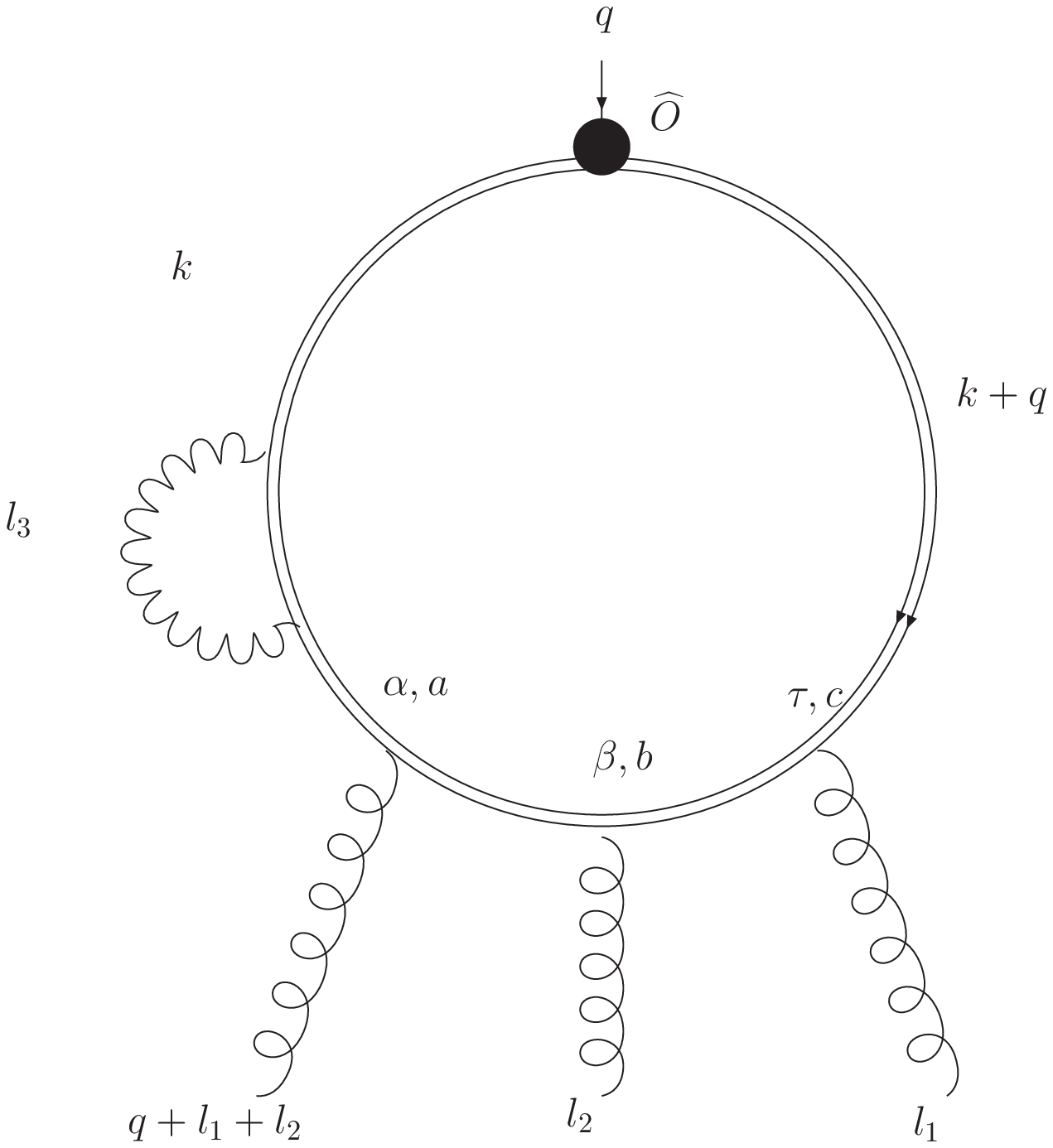}}
\subcaptionbox{2 loop: type V\label{2loop3gluonhigherorder}}
  [.2\linewidth]{\includegraphics[scale=.15]{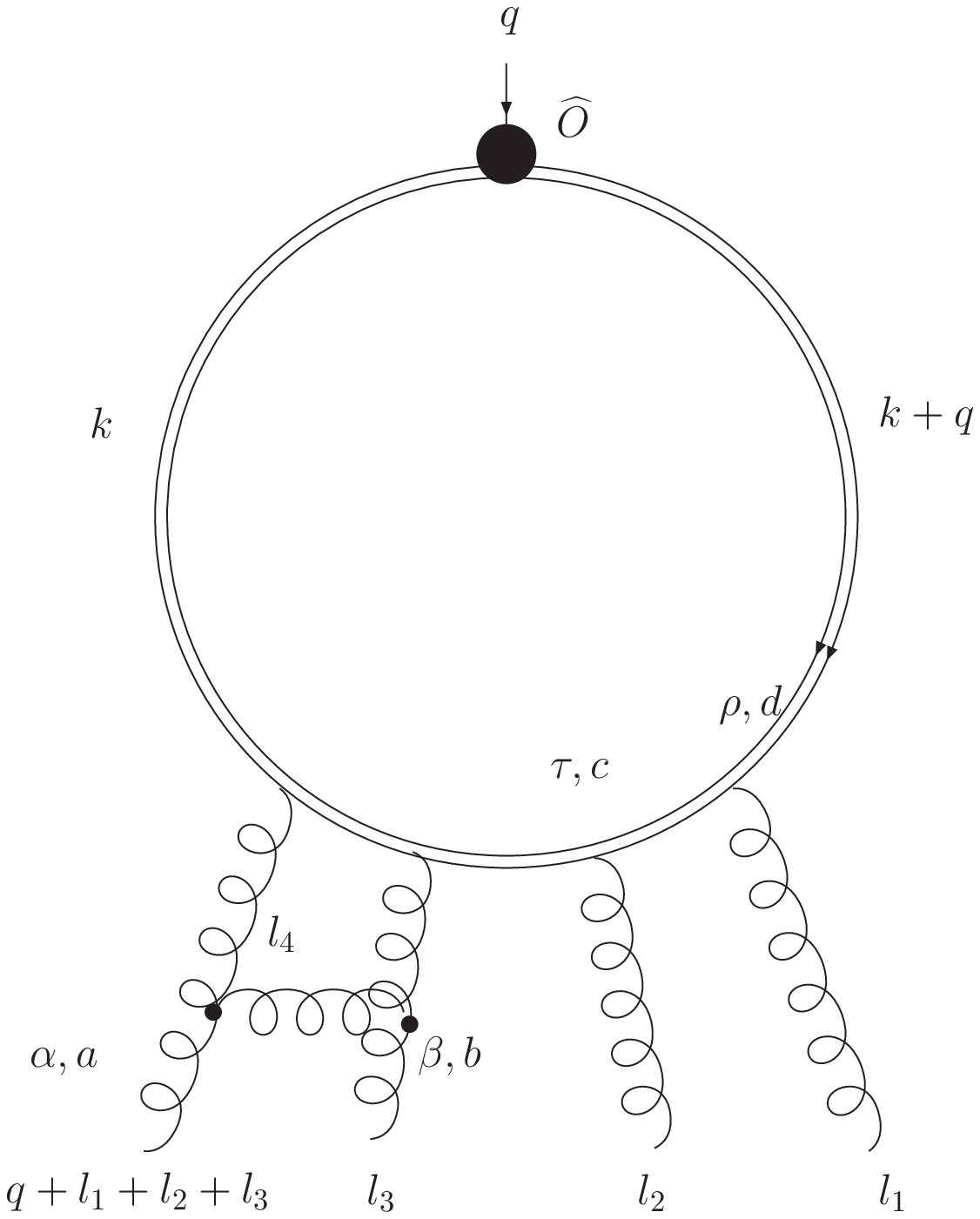}}
\subcaptionbox{2 loop: type VI\label{2loop4gluonhigherorder1}}
  [.2\linewidth]{\includegraphics[scale=.15]{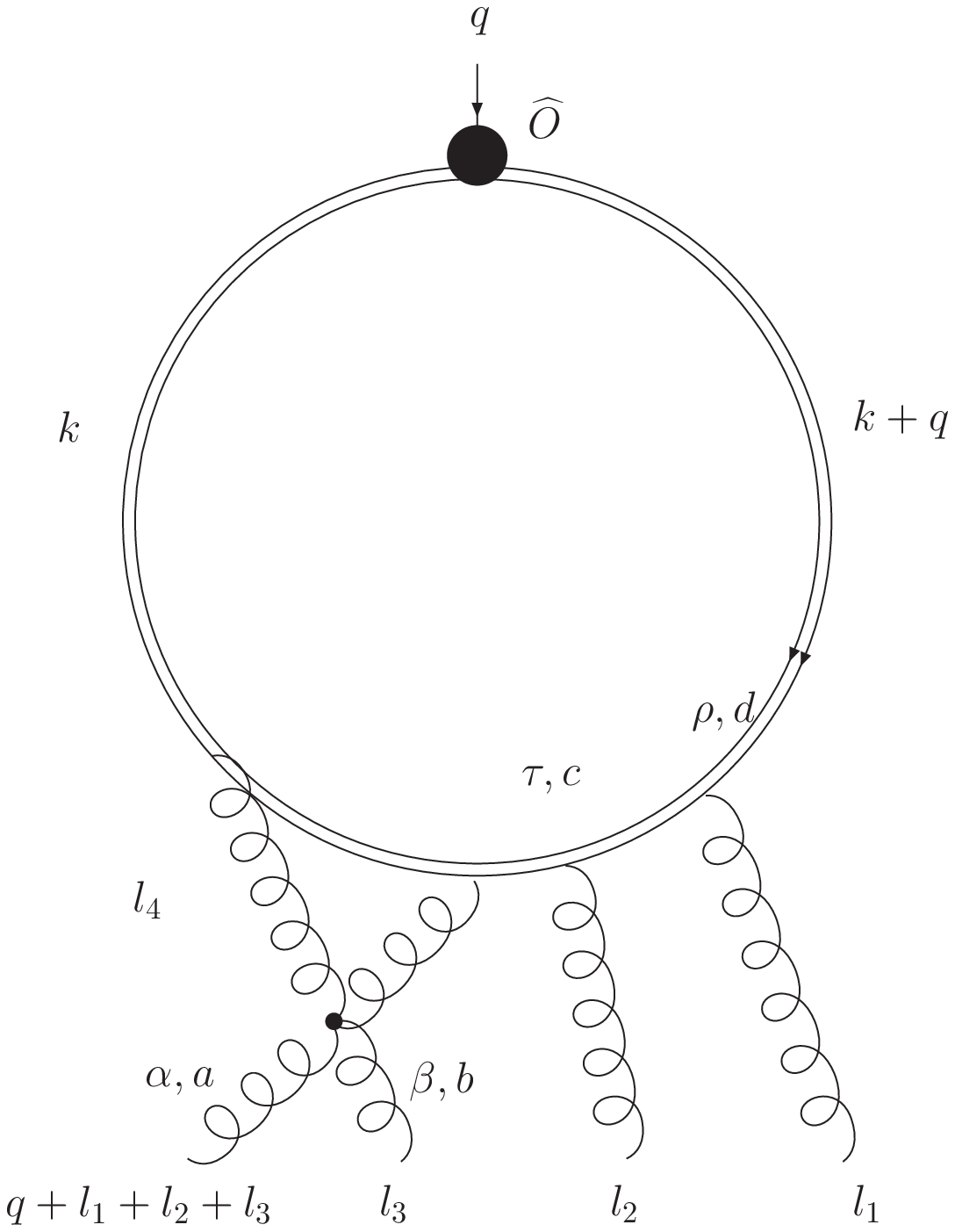}}
\subcaptionbox{2 loop: type VII\label{2loop4gluonhigherorder2}}
  [.2\linewidth]{\includegraphics[scale=.15]{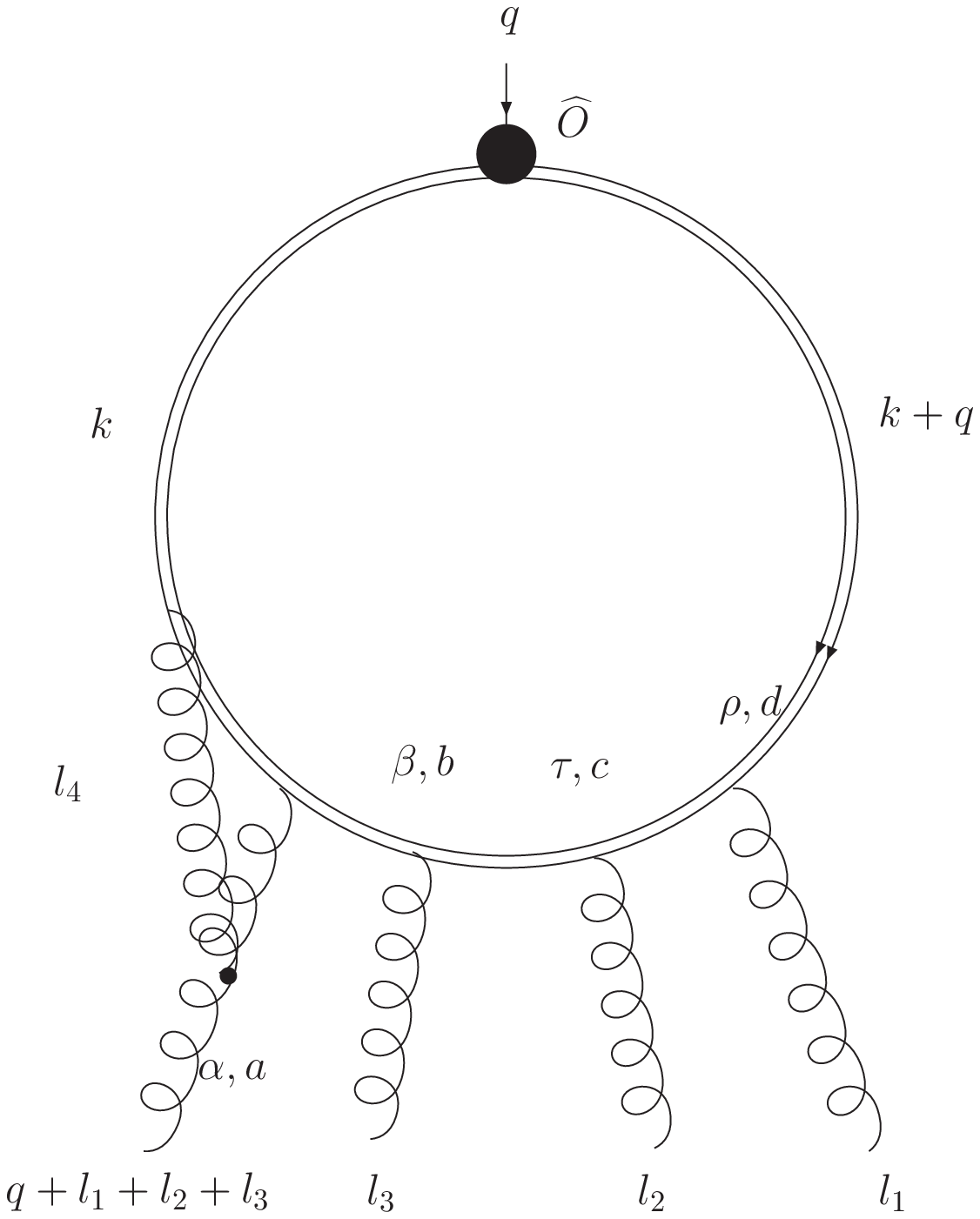}}
\subcaptionbox{2 loop: type VIII\label{2loop5gluonhigherorder}}
  [.2\linewidth]{\includegraphics[scale=.15]{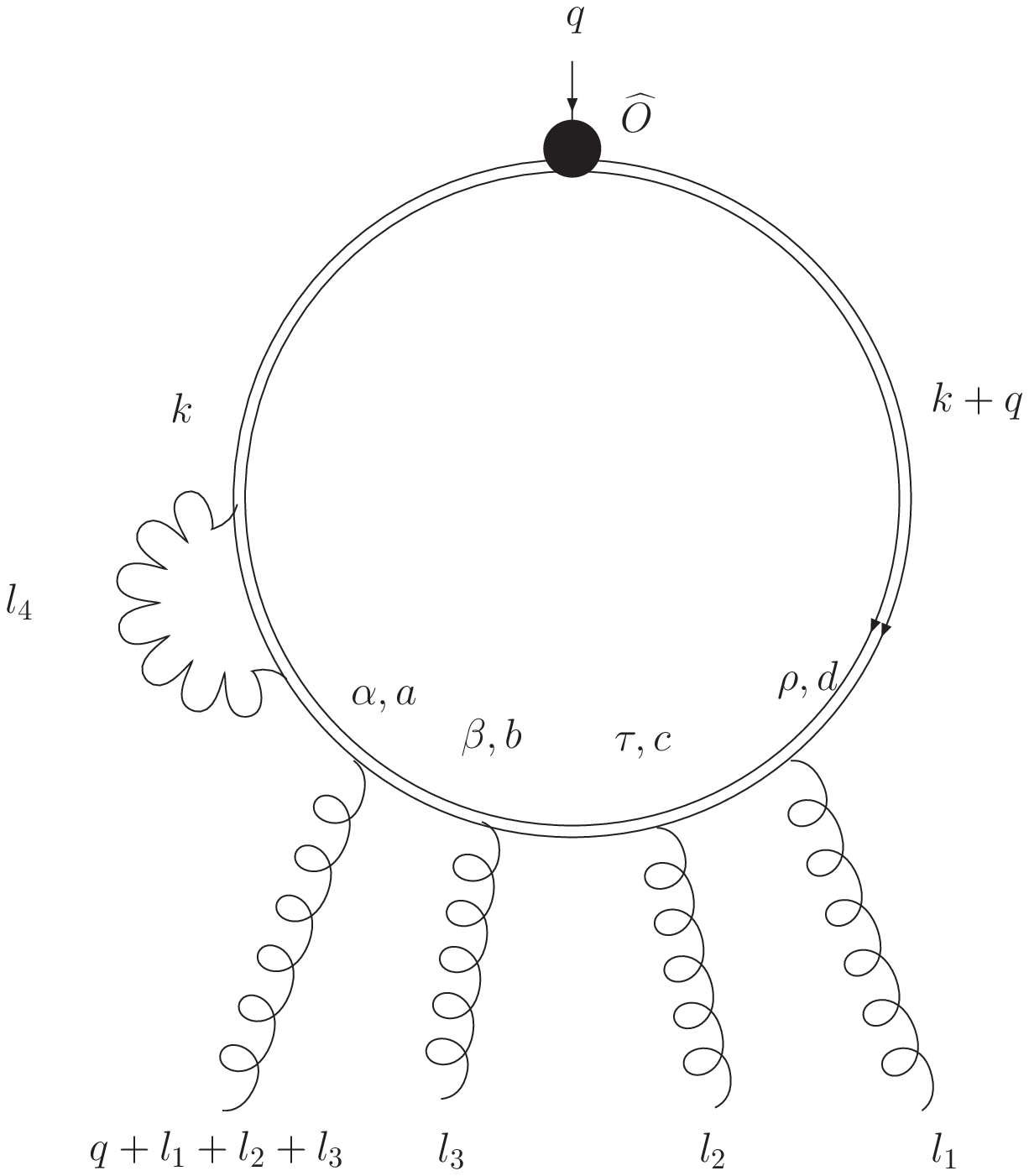}}
\subcaptionbox{3 loop: type I\label{3loopfermion}}
  [.2\linewidth]{\includegraphics[scale=.15]{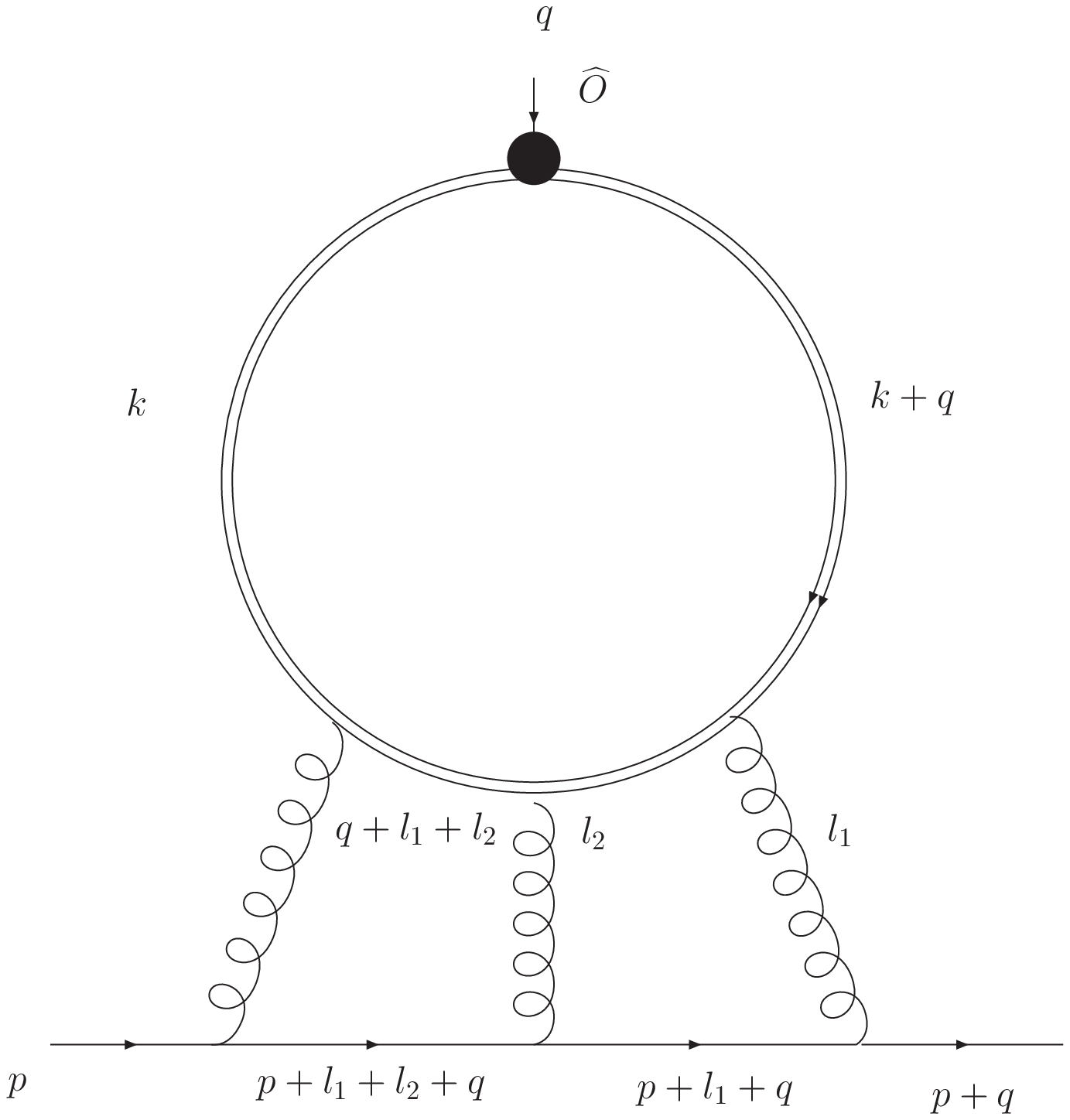}}
\caption{AI-diagrams for the HQME of vector current and tensor current: The blob denotes the operator insertion, the double lines denote heavy quarks}
\label{fig:AI-diagrams}
\end{figure}

Now, we list the AI-diagrams relevant for us as shown in figure \ref{fig:AI-diagrams}.  
We start with one loop diagrams.
Following the arguments given above, the diagrams with three external gluons 
[\ref{1loop}] give the first contributions for the vector current and the tensor current. For the vector current these diagrams have  $\overline{\omega_\gamma}=-4$, while for 
the tensor current they have $\overline{\omega_\gamma}=-3$. Actually these diagrams can be excluded from the computations at the order we consider.
They are  nevertheless, because we want to compare the results  to the HQME-results given in \cite{vc2009}.
The other group of diagrams that is relevant at the order considered in our expansion is that with four external legs [\ref{1loop2}]. By power counting,
these diagrams have  $\overline{\omega_\gamma}=-2$ for the vector current and  $\overline{\omega_\gamma}=-1$ for the tensor current. 
However, our explicit computation and the results of section \ref{sec:loop_structure}  showed that there are no $1/M^2$ or $1/M$ contributions.
This is consistent with the HQME results which predict that all one loop contributions are suppressed up to the order $1/M^4$ for the vector current and $1/M^3$ for the tensor current.\\
Now, consider two-loop  diagrams. Since at least three gluon lines must couple to the fermion loop, the 
AI-diagrams that are relevant for us are given by the figures \ref{2loop3gluon},  \ref{2loop}, \ref{2loop4gluon} and \ref{2loop5gluon} at order  $g_s^5$ and by \ref{2loop3gluonhigherorder}
 \ref{2loop4gluonhigherorder1}, \ref{2loop4gluonhigherorder2}
  and \ref{2loop5gluonhigherorder} at order $g_s^6$.\\ 
The three loop diagrams are the first which contain external  external fermions. In the computation we will only consider one light quark flavor since all light flavors will, up to trivial modifications, yield the same result.
If only two of the gluons are coupled to the fermion line, we have at least $\overline{\omega_\gamma}=-3$ for the vector current and  $\overline{\omega_\gamma}=-2$ for 
the tensor current. Thus, the only possibility is to couple all three gluon legs to the fermion line, see Fig.~ \ref{3loopfermion}. 
For external ghost fields charge conjugation invariance shows the absence of contributions with two external ghosts (due to the ghost number conservation the number of ghosts fields must be even):
\begin{align*}
G_{2}^{a \,b}\left(x,p_1,p_2\right)\quad&=\quad \bra{0}T\mathcal{O}(x)\overline{c}_a\left(p_1\right) c_b\left(p_2\right)\ket{0}\\
                                 \quad&=\quad \frac{\delta^{a\,b}}{8}\bra{0}T\mathcal{O}(x)\overline{c}_d\left(p_1\right) c_d\left(p_2\right)\ket{0}\\
                                  \quad&\xrightarrow{C}\quad (-1)\frac{\delta^{a\,b}}{8}\bra{0}T\mathcal{O}(x)\overline{c}_d\left(p_1\right) c_d\left(p_2\right)\ket{0}\\
                                     \quad&=\quad 0\textnormal{.}
\end{align*}
All nonzero diagrams with more external ghost fields are of higher order than $1/M^2$ in the expansion.\\
Note that to all diagrams given in Fig.~\ref{fig:AI-diagrams} one also has to compute the crossed diagrams.  
\par
There is one additional restriction for the purely gluonic operators in the expansion: The expansion should respect local gauge invariance. Thus, it should contain only 
field strength tensors and covariant derivatives of them. Since the behaviour of the field strength tensor under $C$ is the same as that for the vector potential, the arguments given above
show that the result for both operators must have the schematic structure $tr_c \left(\acom[F_1,F_2]F_3\right)$. For the vector current, current conservation requires that the result must be the 
derivative of some antisymmetric tensor, giving   $\partial \; tr_c \left(\acom[F_1,F_2]F_3\right)$. From power counting it is
now clear that the gluonic contributions to the expansion should
be of order $1/M^4$ for the vector current and $1/M^3$ for the tensor current to all orders of perturbation theory. Our
computations up to $\mathcal{O}(\alpha_s^3)$  confirm these results.

\section{Computation of the AI-diagrams}
We can now use the relation (\ref{eq:AE_general})  to compute the large mass expansion for the diagrams shown in the previous section.
There is one additional simplification.
For the expansion considered here single diagrams may have divergences, but if one takes the sum over all diagrams that contribute
at a given order in the heavy mass and the coupling constant, the divergences cancel. Actually even the sums of all diagrams of a given type (as indicated in Fig.~\ref{fig:AI-diagrams})
are all finite. Thus, it is not necessary to perform the R-operation for them.

To obtain the desired expansion, each AI-diagram has to  be expanded in its small masses and external momenta. However, it is more convenient to rescale the integration momenta according to
$l\rightarrow M l$ (note that this makes the integration variables dimensionless). Since the integrand $I(M,\{l\},\{m\},\{p\})$ is a homogeneous function (here 
$\{l\}$ labels the set of integration momenta and $\{p\}$ external momenta while $\{m\}$ is the set of light masses), we have for an $n$-loop integral 
\begin{align*}
I(M,\{Ml\},\{m\},\{p\})=M^{\omega_\gamma-d n} \overline{I}(M,\{l\},\{m/M\},\{p/M\})\textnormal{.}
\end{align*}
This makes it obvious that we can also expand in $1/M$.\par
After the Taylor expansion of the  of the AI diagrams we obtain several thousand of different tensor integrals and after tensor reduction there are thousands of scalar integrals. In 
the following we explain how the tensor reduction was done and how each type of diagram was computed. The computation was done by a Mathematica program which made use of the package 
`FeynCalc' \cite{Mertig:1990an}. 

In most cases the expansion of the AI-diagrams given above does not result in scalar integrals. This is either because of uncontracted Lorentz-indices in the diagram or because
of scalar products of external and loop momenta occurring in the numerator in the $1/M$-expansion of the integrand. We will explain how the tensor integrals
were reduced to scalar ones. Since the expansion is in all external momenta, the (tensor) integrals that we obtain effectively correspond to diagrams without external momenta; they
are vacuum diagrams. For these, the tensor decomposition is easy: Only the metric tensor is available to represent the tensor structure of a diagram. Thus, an arbitrary integral $F^{\mu_1\dots \mu_{2n}}$ 
(all integrals with an odd number of indices vanish) occurring in the expansion can be written as
\begin{align}
F^{\mu_1\dots \mu_{2n}}=A_1 g^{\mu_1\,\mu_2}\cdot\dots\cdot g^{\mu_{2n-1}\,\mu_{2n}}+\text{other contractions}\textnormal{.}
\label{tensor}
\end{align} 
Here $A_i$ are Lorentz scalars. In general, there are $\frac{(2n)!}{n!\,2^n}=(2n-1)!!$ different contractions. In many cases, the integral is symmetric under the interchange of some of its indices. 
Then one can symmetrize both sides of Eq.~ (\ref{tensor}) such that the number of independent constants $A_I$ is further reduced. Now, contracting  \ref{tensor} with each of the independent 
tensor structures in turn, one obtains a system of equations for the $A_I$ with scalar integrals on the left hand site. Solving for the
constants and reinserting the solution on the right hand side of \ref{tensor},
one obtains a decomposition of the integral in terms of scalar ones.\\
The computation of the scalar integrals is described for each type of AI integral separately. Note however, that the Feynman
rules for the composite operator $\mathcal{O}(x)$ contain an additional factor
of   $e^{i q\cdot x}$ in comparison to the usual Feynman rules for an identical term in the Lagrangian (this results from
the fact that there is no integration over $x$).
\par
The translation from the results of the Feynman diagrams to the operators occurring in the OPE can be done with different sophistication (for example with the help
of the counterterm technique \cite{Anikin:1978tj} or the method of projectors \cite{Gorishnii:1986gn}). Here we simply give operators that yield vertices with the correct external fields and with vertex 
factors that are those polynomials resulting from the computation of the AI-graphs.

The vertex factor corresponding to an operator  $O(x)$ is:
\begin{align*}
\prod_{i=1}^{n_{B}}\frac{\delta}{\delta A_{\mu_{i}}\left(p_{i}\right)}\prod_{j=1}^{n_{F}^{out}}\frac{\delta}{\delta\overline{\Psi}\left(q_{j}\right)}\prod_{l=1}^{n_{B}^{in}}
\frac{\delta}{\delta\Psi\left(k_{l}\right)}\quad O(x)\textnormal{.}
\end{align*}  
It is useful to note that the operator $G^b_{\alpha \beta}$ corresponds to the projection operator on the transversal
component of the momentum that enters the vertex:
\begin{align*}
\frac{\delta}{\delta A^a_{\mu}\left(p\right)}G^b_{\alpha \beta}(x)
\quad=\quad&\frac{\delta}{\delta A^a_{\mu}\left(p\right)} \left\{\partial_{\alpha}\int d^dk\;e^{-i k\cdot x}A^b_{\beta}(k)
                                                 -\partial_{\beta}\int d^dk\;e^{-i k\cdot x}A^b_{\alpha}(k)\;+\;\mathcal{O}\left({g_s}\right)\right\}\\
=\quad & -ie^{-i p\cdot x}\delta^{ab} \left(p_{\alpha}g_{\mu \beta}-p_{\beta}g_{\mu \alpha}\right)\;+\;\mathcal{O}\left({g_s}\right)\textnormal{.}
\end{align*}  
The generalization to higher powers of gluon fields is straightforward.

Additional momentum factors are of course generated by additional derivatives of the gluon field operators. The situation is especially easy for additional factors of
the overall momentum flowing into the vertex:
\begin{align*}
&\prod_{i=1}^{n}\left(\frac{\delta}{\delta A^a_{\mu_i}\left(p_i\right)}\right)\partial^{\nu}\prod_{j=1}^n \left(G^{b_j}_{\alpha_j \beta_j}(x)\right)\\
=\quad & (-i)^{n+1}e^{-i p\cdot x}\;\left(\sum_{i=1}^n p_i^{\nu}\right)\;\left\{\prod_{i=1}^n\left(\delta^{a_i b_{\sigma(i)}} \left(p_{\alpha_{\sigma(i)}}g_{\mu_i \beta_{\sigma(i)}}
 -p_{\beta_{\sigma_i}}g_{\mu_i \alpha_{\sigma(i)}}\right)\;+\;\mathcal{O}\left({g_s}\right)
\right)\right\}\textnormal{.}
\end{align*}  
The $\mathcal{O}(g_s)$-contribution is of higher order in comparison to the diagrams considered here and can be neglected. We shall denote the OPE as
\begin{align*}
\qb \gamma^\mu Q \quad&\overset{M\to \infty}{\simeq}\quad O_{1 loop}^{VC}\;+\;O_{2 loop}^{VC}\;+\;O_{3 loop}^{VC}\\ 
\qb \sigma^{\mu\nu} Q \quad&\overset{M\to \infty}{\simeq}\quad O_{1 loop}^{TC}\;+\;O_{2 loop}^{TC}\;+\;O_{3 loop}^{TC}\textnormal{.}
\end{align*}
Here `$\simeq$' indicates that the relation gives an asymptotic series.

\subsection{One-loop contribution}
\subsubsection{Structure of the fermion loop}\label{sec:loop_structure}

\begin{figure}[h]
\includegraphics[scale=.3]{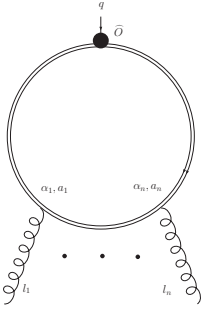}
\caption{Fermion loop with n gluon legs}
\label{fig:nloop}
\end{figure}

The graph \ref{fig:nloop} is the LO-contribution to the matrix element:
\begin{align*}
& G_{a_{1},\dots,a_{n}}^{\alpha_{1}\dots\alpha_{n}}\left(x,x_{1},\dots,x_{n}\right) 
\quad =\quad \left\langle 0\right|T\;\mathcal{O}\left(x\right)\; J_{a_{1}}^{\alpha_{1}}\left(x_{1}\right)\dots 
  J_{a_{n}}^{\alpha_{n}}\left(x_{n}\right)\left|0\right>\textnormal{.}
\end{align*}
Here $J_{a}^{\alpha}\left(x\right)=\left(\overline{Q}\;\gamma^{\alpha}T_{a}Q\right)\left(x\right)$
is the color current. Now, in QCD the color current is not exactly
conserved, but we have $\partial_{\alpha}J_{a}^{\alpha}\left(x\right)=0+\mathcal{O}\left(g_s\right)$.
The fermion loop that we consider contains only the contributions
of LO in $g_s$. Thus, to analyze the loop we can assume that the current
is conserved. Taking the divergence of the matrix element, LO terms
are only generated by moving the derivative inside the time ordering:
\begin{align*}
&\partial_{\alpha_{1}}G_{a_{1},\dots,a_{n}}^{\alpha_{1}\dots\alpha_{n}}\left(x,x_{1},\dots,x_{n}\right)\\  
=\quad&\sum_{i=2}^{n}\delta\left(x_{1}^{0}-x_{i}^{0}\right)\left\langle 0\right||T\;\mathcal{O}\left(x\right)
\;\left[J_{a_{1}}^{0}\left(x_{1}\right)\;,\; J_{a_{i}}^{\alpha_{i}}\left(x_{i}\right)\right]\prod_{j=2,j\neq i}^{n}J_{a_{j}}^{\alpha_{j}}
   \left(x_{j}\right)\left|0\right>\\
  &+\delta\left(x_{1}^{0}-x_{i}^{0}\right)\left\langle 0\right|T\;\left[J_{a_{1}}^{0}\left(x_{1}\right)\;,
\;\mathcal{O}\left(x\right)\right]\prod_{j=2}^{n}J_{a_{j}}^{\alpha_{j}}\left(x_{j}\right)\left|0\right\rangle\textnormal{.}
 \end{align*}
Now, the commutators are determined by the transformation properties
of the operators under $SU(3)_{c}$-transformations:
\begin{align*}
&\left[J_{a_{1}}^{0}\left(x_{1}\right)\;,\;\mathcal{O}\left(x\right)\right]  &=\;&0\\
&\left[J_{a_{1}}^{0}\left(x_{1}\right)\;,\; J_{a_{i}}^{\alpha_{i}}\left(x_{i}\right)\right]  &=\;& if_{a_{1}a_{i}b}\; J_{b}^{\alpha_{i}}\left(x_{i}\right)
\;\delta^{(3)}\left(\vec{x_{1}}-\vec{x_{i}}\right)\textnormal{.}
\intertext{Here we assume that possible Schwinger-terms can be neglected for vacuum expectation values that we are computing.
This gives the result} 
&\partial_{\alpha_{1}}G_{a_{1},\dots,a_{n}}^{\alpha_{1}\dots\alpha_{n}}\left(x_{1},\dots,x_{n}\right) &=\;&i\sum_{i=2}^{n}\delta^{\left(d\right)}\left(x_{1}-x_{i}\right) f_{a_{1}a_{i}b}\left\langle 0\right|T\mathcal{O}\left(x\right) 
J_{b}^{\alpha_{i}}\left(x_{i}\right)\prod_{j=2,j\neq i}^{n}J_{a_{j}}^{\alpha_{j}}\left(x_{j}\right)\left|0\right\rangle \textnormal{.}
\intertext{In momentum space we have} 
&\left(p_{1}\right)_{\alpha_{1}}G_{a_{1},\dots,a_{n}}^{\alpha_{1}\dots\alpha_{n}}\left(x,p_{1},\dots,p_{n}\right) 
&=\;& \sum_{i=2}^{n}\; f_{a_{1}a_{i}b}\;\left\langle 0\right|T\;\mathcal{O}\left(x\right)\; J_{b}^{\alpha_{i}}\left(p_{1}+p_{i}\right)\prod_{j=2,j\neq i}^{n}
J_{a_{j}}^{\alpha_{j}}\left(p_{j}\right)\left|0\right\rangle \textnormal{.}
\end{align*}
Obviously, there are similar relations for all external momenta $p_{2}\dots p_{n}$.
Now,we can draw conclusions for the case of three and four external gluons.\\
For 3 external gluons, we have
\begin{align*}
\left(p_{1}\right)_{\alpha_{1}}G_{a_{1},\dots,a_{n}}^{\alpha_{1}\dots\alpha_{n}}\left(x,p_{1},p_{2},p_{3}\right)  \quad=\quad& f_{a_{1}a_{2}b}\;\left\langle 0\right|T\mathcal{O}\left(x\right) J_{b}^{\alpha_{2}}\left(p_{1}+p_{2}\right)J_{a_{3}}^{\alpha_{3}}\left(p_{3}\right)\left|0\right\rangle \\
 &+ f_{a_{1}a_{3}b}\;\left\langle 0\right|T\;\mathcal{O}\left(x\right)\; J_{b}^{\alpha_{3}}\left(p_{1}+p_{3}\right)J_{a_{2}}^{\alpha_{2}}\left(p_{3}\right)\left|0\right\rangle \\
  =\quad &0\textnormal{.}
\end{align*}
The last equality follows from $C$-invariance as was discussed above.
From this relation (and from the corresponding ones for $p_{2},p_{3}$)
it is clear that 
\begin{align*}
G_{a_{1},\dots,a_{n}}^{\alpha_{1}\dots\alpha_{n}}\left(x,p_{1},p_{2},p_{3}\right)  \quad=\quad &\prod_{i=1}^{3}\left(p_{i}^{\mu_{i}}g^{\nu_{i}\alpha_{i}}-p_{i}^{\nu_{i}}g^{\mu_{i}\alpha_{1}}\right)
\left(F_{a_{1},\dots,a_{n}}\right)_{\mu_{1}\dots\mu_{n},\nu_{1},\dots,\nu_{n}}\left(x,p_{1},p_{2},p_{3}\right)\textnormal{.}
\end{align*}

Now, the power counting arguments given above apply. These show that
the matrix element for the tensor current is suppressed as $M^{-3}$
and the vector current as $M^{-4}$.\\
 For the case of four external gluons we have
\begin{align*}
\left(p_{1}\right)_{\alpha_{1}}G_{a_{1},\dots,a_{n}}^{\alpha_{1}\dots\alpha_{n}}\left(x,p_{1},p_{2},p_{3},p_{4}\right)  \quad=\quad& f_{a_{1}a_{2}b}\;\left\langle 0\right|T\;\mathcal{O}\left(x\right)\; J_{b}^{\alpha_{2}}\left(p_{1}+p_{2}\right)J_{a_{3}}^{\alpha_{3}}\left(p_{3}\right)J_{a_{4}}^{\alpha_{4}}\left(p_{4}\right)\left|0\right\rangle \\
 &+ f_{a_{1}a_{3}b}\;\left\langle 0\right|T\;\mathcal{O}\left(x\right)\; J_{b}^{\alpha_{3}}\left(p_{1}+p_{3}\right)J_{a_{2}}^{\alpha_{2}}\left(p_{3}\right)J_{a_{4}}^{\alpha_{4}}\left(p_{4}\right)\left|0\right\rangle \\
 &+ f_{a_{1}a_{4}b}\;\left\langle 0\right|T\;\mathcal{O}\left(x\right)\; J_{b}^{\alpha_{4}}\left(p_{1}+p_{4}\right)J_{a_{2}}^{\alpha_{2}}\left(p_{3}\right)J_{a_{3}}^{\alpha_{3}}\left(p_{3}\right)\left|0\right\rangle \\
 =\quad & \mathcal{O}\left(\frac{1}{M^{n}}\right)\textnormal{.}\end{align*}
Here $n=3$ for the tensor current and $n=4$ for the vector current. Since
this applies for all external momenta, those parts of the matrix elements
that are not suppressed as described above must have at least four
projection operators. From power counting and the general structure
of the expansion for tensor current and vector current we see that
the contributions with these projectors are suppressed by $M^{-5}$and
$M^{-6}$, respectively.

\subsubsection{Feynman integrals for the for the one-loop contribution}
The only 1-loop integrals to compute  are shown in \ref{1loop} and \ref{1loop2}. The integrals were computed for a fixed assignment of external
momenta. Afterwards the permutations in external momenta were done to generate the crossed diagrams. 
For the diagram  \ref{1loop} we have the analytic expression:
\begin{align*}
 I^{I}_{1\;loop}\quad=&\quad -\left(i g_s\right)^3\;tr_c \left(T_a T_b T_c\right)
\quad\int \frac{d^d k}{\left( 2 \pi\right)^d}\;
tr_\gamma \bigg(\Gamma\; \frac{1}{\slashed{k}-M} \;\gamma^{\alpha}\;\frac{1}{\slashed{k}+\slashed{q}+\slashed{l_1}+\slashed{l_2}-M}\\
&\quad \gamma^{\beta}\;\frac{1}{\slashed{k}+\slashed{q}+\slashed{l_1}-M}\;\gamma^{\tau}\;\frac{1}{\slashed{k}+\slashed{q}-M}\bigg)\textnormal{.}
\intertext{Here we use $\Gamma=\gamma^\mu$ for the vector current and $\Gamma=\sigma^{\mu \nu}$ for the tensor current.
For the diagram \ref{1loop2} with four external lines we have}
 I^{II}_{1\;loop}\quad=&\quad -i\left(i g_s\right)^4\;tr_c \left(T_a T_b T_c T_c\right)
\quad\int \frac{d^d k}{\left( 2 \pi\right)^d}\;
tr_\gamma \bigg(\Gamma\; \frac{1}{\slashed{k}-M} \;\gamma^{\alpha}\;\frac{1}{\slashed{k}+\slashed{q}+\slashed{l_1}+\slashed{l_2}+\slashed{l_3}-M}\\
&\quad \gamma^{\beta}\;\frac{1}{\slashed{k}+\slashed{q}+\slashed{l_1}+\slashed{l_2}-M}\;\gamma^{\tau}\;\frac{1}{\slashed{k}+\slashed{l_1}+\slashed{q}-M}
\;\gamma^{\rho}\;\frac{1}{\slashed{k}+\slashed{q}-M}\bigg)\textnormal{.}
\end{align*}
The master integrals that occur after the expansion are:
\begin{align*}
F(n)\quad &=\quad \int \frac{d^d k}{\left( 2 \pi\right)^d}\;\left(\frac{1}{k^2-1}\right)^n
 \quad =\quad\frac{(-1)^{n}i}{\left(4\pi\right)^{d/2}}\quad\frac{\Gamma\left(n-\frac{d}{2}\right)}{\Gamma\left(n\right)}\textnormal{.}
\end{align*}
We use $l_3=-q-l_1-l_2$ to simplify the notation (this is the momentum entering in the last gluon vertex). \\

\subsubsection{Result for the vector current}
First consider the vector current in \ref{1loop}. The result of expanding these diagrams up to the order $1/M^4$ reads:
\begin{alignat*}{5}
&\left(I_{VC}\right)^{I}_{1\;loop}\quad +&&\textnormal{perm.}\\ 
=&\;\frac{g_s^3 d_{abc}\;e^{i q\cdot x}}{144 M^4 \pi ^2}\bigg\{
&\frac{7}{5}\bigg[& \left(g^{\beta \rho} l_2^{\gamma}-g^{\beta \gamma} l_2^{\rho}\right)
   \left(g^{\mu}_{\gamma} q^{\omega}-g^{\mu \omega} q_{\gamma}\right)  \left(g^{\alpha}_{\omega} l_3^{\delta}-g^{\alpha \delta} \left(l_3\right)_{\omega}\right)
   \left(\left(l_1\right)_{\rho} g^{\tau }_{\delta}-g^{\tau }_{\rho}\left(l_1\right)_{\delta}\right)\\
&&&+ \left(g^{\beta \rho} l_2^{\gamma}-g^{\beta \gamma} l_2^{\rho}\right) \left(g^{\alpha}_{\gamma}
   l_3^{\omega}-g^{\alpha \omega} \left(l_3\right)_{\gamma}\right)  \left(g^{\mu}_{\omega} q^{\delta}-g^{\mu \delta} q_{\omega}\right)
   \left(\left(l_1\right)_{\rho} g^{\tau }_{\delta}-g^{\tau }_{\delta}\left(l_1\right)_{\delta}\right)\\
&&&+ \left(g^{\alpha \rho} l_3^{\gamma}-g^{\alpha \gamma}
    l_3^{\rho}\right) \left(g^{\beta}_{\gamma} l_2^{\omega}-g^{\beta \omega} \left(l_2\right)_{\gamma}\right)  \left(g^{\mu}_{\omega} q^{\delta}-g^{\mu \delta} q_{\omega}\right)
    \left(\left(l_1\right)_{\rho} g^{\tau }_{\delta}-g^{\tau }_{\delta} \left(l_1\right)_{\delta}\right)\bigg]\\
&&-\frac{1}{2}\bigg[&
   \left(l_1^{\rho} g^{\gamma\tau }-g^{\rho\tau } l_1^{\gamma}\right) \left(g^{\mu}_{\rho} q_{\gamma}-g^{\mu}_{\gamma} 
   q_{\rho}\right) \left(g^{\beta \delta} l_2^{\omega}-g^{\beta \omega} l_2^{\delta}\right) \left(g^{\alpha}_{\omega}
   \left(l_3\right)_{\delta}-g^{\alpha}_{\delta} \left(l_3\right)_{\omega}\right)\\
&&&+ \left(l_1^{\rho}
   g^{\gamma\tau }-g^{\rho\tau } l_1^{\gamma}\right) \left(g^{\alpha}_{\rho} \left(l_3\right)_{\gamma}-g^{\alpha}_{\gamma}
   \left(l_3\right)_{\rho}\right) \left(g^{\beta \delta} l_2^{\omega}-g^{\beta \omega} l_2^{\delta}\right) \left(g^{\mu}_{\omega}
   q_{\delta}-g^{\mu}_{\delta} q_{\omega}\right)\\
&&&+\left(l_1^{\rho} g^{\gamma\tau }-g^{\rho\tau }
   l_1^{\gamma}\right) \left(g^{\beta}_{ \rho} \left(l_2\right)_{\gamma}-g^{\beta}_{ \gamma} \left(l_2\right)_{\rho}\right) \left(g^{\alpha \delta}
   l_3^{\omega}-g^{\alpha \omega} l_3^{\delta}\right) \left(g^{\mu}_{ \omega} q_{\delta}-g^{\mu}_{ \delta} q_{\omega}\right)
   \bigg]\bigg\}\textnormal{.}
\end{alignat*}
This way to write the result clearly shows the expected projector structure of the result: Contracting with $q_\mu$, $(l_1)_\alpha$, $(l_2)_\beta$ or $(l_3)_\tau$ we 
obtain zero. \par
Due to the projector-structure of the results it is easy to write down the corresponding operators. After a short calculation, we have, omitting the argument $x$ and neglecting higher 
terms in the coupling constant:
\begin{align*}
&\left(I_{VC}\right)^{I}_{1\;loop}\quad +\textnormal{perm.}\\
=\;&\A[\alpha,a,l_3]\A[\beta,b,l_2]\A[\tau,c,l_1]\,\bigg\{\frac{-g_s^3 d_{a1\,a2\,a3}}{144 M^4 \pi ^2}\;
\partial_{\rho}\left(
-\frac{7}{5}G_{a1}^{\rho\delta }(x)\left(G_{a2}\right)_{\delta\gamma }G_{a3}^{\gamma\mu }
+\frac{1}{2}G_{a1}^{\mu \rho}(x)\left(G_{a2}\right)_{\delta\gamma }G_{a3}^{\delta\gamma } \right)\bigg\}\\
=\;&\A[\alpha,a,l_3]\A[\beta,b,l_2]\A[\tau,c,l_1]\,\bigg\{\frac{-g_s^3}{72 M^4 \pi ^2}\;tr_c\bigg[
\nabla_{\rho}\,,\,
\frac{7}{5}\acom[G^{\mu \gamma},G_{ \gamma \delta}]G^{\delta \rho }+G^{\delta \gamma}G_{\delta\gamma } G^{\mu \rho}\bigg]\bigg\}\textnormal{.}
\end{align*}
With the application to instanton solutions in mind, we introduce the field strength tensor $F_{\mu\nu}=g_s G_{\mu\nu}$.
Thus, the contribution can be generated by
\begin{align}
O^{VC}_{1 loop}\quad=\quad\frac{-1}{72 M^4 \pi ^2}\;tr_c\bigg[
\nabla_{\rho}\,,\,\frac{7}{5}\acom[F^{\mu \gamma},F_{ \gamma \delta}]F^{\delta \rho }+F_{\delta \gamma}F^{\delta\gamma } F^{\mu \rho}\bigg]\bigg\}\;+\;\mathcal{O}\left(\frac{1}{M^6}\right)\textnormal{.}
\label{eq:as_vc_1loop}
\end{align}
This is exactly the result from the calculation of \cite{vc2009} after continuing to Minkowski-spacetime and rescaling the gluon fields.\\
The contribution of figure \ref{1loop2} will not be given here due to its huge size. However, it corresponds 
exactly to the $\mathcal{O}\left(g_S^4 \right)$-contribution to Eq.~ (\ref{eq:as_vc_1loop}).

\subsubsection{Tensor current}
For the tensor current a similar computation gives the result:
\begin{align*}
\left(I_{TC}\right)^{I}_{1\;loop} +\textnormal{perm.}&\\
=\quad\frac{-i g_s^3 d_{abc}\;e^{i q\cdot x}}{48 M^3 \pi ^2}\bigg\{\quad&
 \bigg[\left(g^{\alpha \mu } \left(l_3\right)_{\rho}-g^{\alpha}_{ \rho}   l_3^{\mu }\right) 
         \left(l_1^{\rho} g^{\gamma\tau }-g^{\rho\tau } l_1^{\gamma}\right) 
         \left(g^{\beta}_{ \gamma} l_2^{\nu }-g^{\beta \nu } \left(l_2\right)_{\gamma}\right) -\bigg(\mu \leftrightarrow \nu\bigg)\bigg]\\ 
+ &\bigg[\left(g^{\alpha \mu } \left(l_3\right)_{\rho}-g^{\alpha}_{ \rho}   l_3^{\mu }\right) 
         \left(l_1^{\nu } g^{\gamma\tau }-g^{\nu \tau } l_1^{\gamma}\right) 
         \left(g^{\beta}_{ \gamma} l_2^{\rho}-g^{\beta \rho} \left(l_2\right)_{\gamma}\right)-\bigg(\mu \leftrightarrow \nu\bigg)\bigg]\\ 
+ &\bigg[\left(g^{\mu \tau } l_1^{\rho}-l_1^{\mu } g^{\rho\tau }\right)   
         \left(g^{\beta}_{ \gamma} l_2^{\nu }-g^{\beta \nu } \left(l_2\right)_{\gamma}\right) 
         \left(g^{\alpha \gamma}   \left(l_3\right)_{\rho}-g^{\alpha}_{ \rho} l_3^{\gamma}\right)-\bigg(\mu \leftrightarrow \nu\bigg)\bigg]\\ 
-&\phantom{\bigg[} \left(g^{\alpha \nu } l_3^{\mu }-g^{\alpha \mu } l_3^{\nu }\right)
  \left(l_1^{\rho} g^{\gamma\tau   }-g^{\rho\tau } l_1^{\gamma}\right) 
  \left(g^{\beta}_{ \rho} \left(l_2\right)_{\gamma}-g^{\beta}_{ \gamma} \left(l_2\right)_{\rho}\right) \\ 
-&\phantom{\bigg[} \left(g^{\beta \nu } l_2^{\mu }-g^{\beta \mu } l_2^{\nu }\right)   
   \left(l_1^{\rho} g^{\gamma\tau }-g^{\rho\tau } l_1^{\gamma}\right) 
   \left(g^{\alpha}_{   \rho} \left(l_3\right)_{\gamma}-g^{\alpha}_{ \gamma} \left(l_3\right)_{\rho}\right) \\ 
-&\phantom{\bigg[} \left(l_1^{\mu } g^{\nu   \tau }-g^{\mu \tau } l_1^{\nu }\right) 
   \left(g^{\beta}_{ \gamma} l_2^{\rho}-g^{\beta \rho}   \left(l_2\right)_{\gamma}\right) 
   \left(g^{\alpha}_{ \rho} l_3^{\gamma}-g^{\alpha \gamma}   \left(l_3\right)_{\rho}\right)\bigg\}\textnormal{.}
\end{align*}
As expected the result vanishes for contractions with $(l_1)_\alpha$, $(l_2)_\beta$ or $(l_3)_\tau$.\\ 
Again, we can rewrite this in terms of an operator insertion:
\begin{align*}
&\left(I_{TC}\right)^{I}_{1\;loop} +\textnormal{perm.}\\
=\;&\A[\alpha,a,l_3]\A[\tau,c,l_1]\A[\beta,b,l_2]\;tr_c\bigg\{\frac{-g_s^3}{24 M^3 \pi ^2} \bigg[
G^{\gamma \nu}G^{\rho \mu}G_{ \gamma \rho}-G^{\gamma \mu}G^{\rho \nu}G_{\gamma \rho}+G^{ \gamma\rho}G_{\gamma \rho}G^{\mu \nu}\bigg]\bigg\}\textnormal{.}
\end{align*}
Thus, we can write this contribution in terms of the operator insertion 
\begin{align}
O^{TC}_{1 loop}\quad=\quad
tr_c\bigg\{\frac{-1}{24 M^3 \pi ^2} \bigg[F^{\gamma \nu}F^{\rho \mu}F_{ \gamma \rho}-F^{\gamma \mu}F^{\rho \nu}F_{\gamma \rho}+F^{ \gamma\rho}F_{\gamma \rho}F^{\mu \nu}\bigg]\bigg\}+\mathcal{O}\left(\frac{1}{M^5}\right) \textnormal{.}
\label{eq:as_tc_1loop}
\end{align}
which coincides with the result of  \cite{Franz:2000ee} if one takes into account the analytic continuation to Minkowski-spacetime and the rescaling of the gluon fields. As in the case of the vector current, the diagram 
figure \ref{1loop2}  corresponds to the $\mathcal{O}\left(g_S^4 \right)$-contribution to equation \ref{eq:as_tc_1loop}.

\subsection{Two-loop contribution}\label{main:2loop}
Since there are several different kinds of integrals at two-loop level, their explicit forms are given in the Appendices \ref{appendix_2loop_I} and \ref{appendix_2loop_II}.
The integrals that have to be computed can be done in closed form and are of the type
\begin{align}
&F(n_1,n_2)\quad=\quad \int\frac{d^{d}k\; d^{d}l_{2}}{\left(2\pi\right)^{2d}}\quad\frac{1}{\left[k^{2}-M^{2}\right]^{n_{1}}\left[\left(k+l_{2}\right)^{2}-M^{2}\right]^{n_{2}}\left[l_{2}^{2}\right]^{n_{3}}}\nonumber\\
         =&\quad   \frac{(-1)^{n_1+n_2+n_3+1}}{\left(4\pi\right)^{d}}\left(\frac{\Gamma\left(n_1+n_2+n_3-d\right)\Gamma\left(n_1+n_3-\frac{d}{2}\right)\Gamma\left(n_2+n_3-\frac{d}{2}\right)\Gamma\left(\frac{d}{2}-n_3\right)}{\Gamma\left(n_1+n_2+2n_3-d\right)\Gamma\left(n_1\right)\Gamma\left(n_2\right)\Gamma\left(\frac{d}{2}\right)}\right)\textnormal{.}
\label{eq:2loop_master_integral}
\end{align}
As was discussed in section \ref{sec:AI-diagrams} there are no gauge invariant purely gluonic operators that can contribute to the order $1/M^{2}$ for the vector current or $1/M^{3}$ for the tensor current. Therefore,
the $1/M^{2}$($1/M$)-contributions of the 2-loop diagrams should add up to $0$ at each order in perturbation theory. At $\mathcal{O}\left(g_s^5\right)$ and $\mathcal{O}\left(g_s^6\right)$
the results given in the Appendices \ref{appendix_2loop_I} and \ref{appendix_2loop_II} confirm this reasoning:
\begin{align*}
O_{2\;loop}^{VC}\quad&=\quad0\;+\;\mathcal{O}\left(\frac{1}{M^4}\right)\\
O_{2\;loop}^{TC}\quad&=\quad0\;+\;\mathcal{O}\left(\frac{1}{M^3}\right)\textnormal{.}
\end{align*}

\subsection{Three-loop contribution}
In this case we have to compute the three loop integrals with external fermions (\ref{3loopfermion}). These are given by 
\begin{align*}
 I_{3\;loop}\quad=&\quad i\left(i g_s\right)^6  \;\left[tr_c \left(T_a T_b T_c\right)\;T_a\;T_b\;T_c\right]\;
\int \frac{d^d k\;d^d l_1\;d^d l_2}{\left( 2 \pi\right)^{3d}}\;\bigg\{
\frac{1}{l_1^2}\;\frac{1}{l_2^2}\;\frac{1}{\left(q+l_1+l_2\right)^2}\\
&\quad
tr_\gamma \bigg(\Gamma\; \frac{1}{\slashed{k}-M} \;\gamma^{\alpha}\;\frac{1}{\slashed{k}+\slashed{q}+\slashed{l_1}+\slashed{l_2}-M}
\;\gamma^{\beta}\;\frac{1}{\slashed{k}+\slashed{q}+\slashed{l_1}-M}
\;\gamma^{\tau}\;\frac{1}{\slashed{k}+\slashed{q}-M}\bigg)\\
&\quad \gamma^{\tau}\;\frac{1}{\slashed{p}+\slashed{q}+\slashed{l_1}-m}\;\gamma^\beta\;\frac{1}{\slashed{p}+\slashed{q}+\slashed{l_1}+\slashed{l_2}-m}\;\gamma^\alpha\;\bigg\}\textnormal{.}
\end{align*}
Here $m$ is the mass of any of the light quark flavors. \\
The integrals that have to be computed are of the form 
\begin{align*}
&I\left(\alpha_{1},\alpha_{2},\alpha_{3},\alpha_{4},\alpha_{5},\alpha_{6}\right)\\=&\int\frac{d^{d}k\, d^{d}l_{1}\, d^{d}l_{2}}{\left(2\pi\right)^{3d}}\frac{1}{\left[k^{2}-1\right]^{\alpha_{1}}\left[\left(k+l_{1}\right)^{2}-1\right]^{\alpha_{2}}\left[\left(k+l_{1}+l_{2}\right)^{2}-1\right]^{\alpha_{3}}\left[l_{1}^{2}\right]^{\alpha_{4}}\left[l_{2}^{2}\right]^{\alpha_{5}}\left[\left(l_{1}+l_{2}\right)^{2}\right]^{\alpha_{6}}}
\end{align*}
These integrals are hard to compute directly, but there is a way of computing them by the integration by parts (IBP) technique. Here we give a slightly modified version of the approach of  \cite{Broadhurst:1991fi}, 
where exactly these kinds of integrals where computed. One should also note that there are several software packages that compute such kinds of integrals (`MATAD' \cite{Steinhauser:2000ry}, 
`FIRE' \cite{Smirnov:2008iw}). However, due to the fact that these
integrals occur as an intermediate step after expanding in the heavy quark mass and tensor reduction of the result and that one has to do at the order of $50.000$ of these integrals 
it was more convenient to write an own program that is specialized on exactly this type of integral.\\

\subsubsection{The IBP-technique}\label{main:3loop}
In order to reduce the number of integrals to compute, it is important to note some of the symmetry properties of these integrals:
\begin{align*}
 & I\left(\alpha_{1},\alpha_{2},\alpha_{3},\alpha_{4},\alpha_{5},\alpha_{6}\right) \quad = \quad    I\left(\alpha_{2},\alpha_{3},\alpha_{1},\alpha_{5},\alpha_{6},\alpha_{4}\right)\\
 & I\left(\alpha_{1},\alpha_{2},\alpha_{3},\alpha_{4},\alpha_{5},\alpha_{6}\right)  \quad =\quad     I\left(\alpha_{1},\alpha_{3},\alpha_{2},\alpha_{6},\alpha_{5},\alpha_{4}\right)\\
 & I\left(\alpha_{1},\alpha_{2},\alpha_{3},\alpha_{4},\alpha_{5},\alpha_{6}\right)  \quad = \quad    I\left(\alpha_{2},\alpha_{1},\alpha_{3},\alpha_{4},\alpha_{6},\alpha_{5}\right)\\
 & I\left(\alpha_{1},\alpha_{2},\alpha_{3},\alpha_{4},\alpha_{5},\alpha_{6}\right)  \quad = \quad    I\left(\alpha_{3},\alpha_{2},\alpha_{1},\alpha_{5},\alpha_{4},\alpha_{6}\right)\\
 & I\left(\alpha_{1},\alpha_{2},\alpha_{3},\alpha_{4},\alpha_{5},\alpha_{6}\right)  \quad = \quad    I\left(\alpha_{3},\alpha_{1},\alpha_{2},\alpha_{6},\alpha_{4},\alpha_{5}\right)\textnormal{.}
\end{align*}
These relations will be used such that the series of indices $\alpha_{1}$,$\alpha_{2}$, $\alpha_{3}$ is always decreasing unless one of the
indices $\alpha_{4}$, $\alpha_{5}$,$\alpha_{6}$ is non-positive and $\alpha_{1}$, $\alpha_{2}$, $\alpha_{3}$ are positive. In these cases the indices $\alpha_{4}$, $\alpha_{5}$,$\alpha_{6}$ will
be chosen to be decreasing.\\
We will now show that we can use the IBP-technique to reduce an integral with arbitrary values of  the indices $\alpha_{i}$ to integrals which fall in one of the following classes: 
\begin{itemize}
 \item[1.] $\alpha_{1}\leq0,\alpha_{2}\leq0$ or  $\alpha_{3}\leq0 $
 \item[2.] $\alpha_{5}\leq0$ and $\alpha_{6}\leq0$ .
\end{itemize}
Afterwards, we show how to compute the integrals for these special cases.\par
Using the relations
\begin{align*}
  k\cdot l_{1}  \quad =&\quad \frac{1}{2}\left[\left(k+l_{1}\right)^{2}-1\right]-\frac{1}{2}\left[k^{2}-1\right]-\frac{1}{2}\left(l_{1}\right)^{2} \sim \quad \frac{1}{2}\left(2^{-}\right)-\frac{1}{2}\left(1^{-}\right)-\frac{1}{2}\left(4^{-}\right)\\
  k\cdot l_{2} \quad \sim & \quad \frac{1}{2}\left(3^{-}\right)-\frac{1}{2}\left(2^{-}\right)+\frac{1}{2}\left(4^{-}\right)-\frac{1}{2}\left(6^{-}\right)\\
  l_{1}\cdot l_{2} \quad \sim& \quad\frac{1}{2}\left(6^{-}\right)-\frac{1}{2}\left(4^{-}\right)-\frac{1}{2}\left(5^{-}\right)\textnormal{.}
\end{align*}
with the standard-notation
\begin{align*}
 & \left(i^{\pm}\right)I\left(\dots,\alpha_{i},\dots\right)  \quad=\quad  I\left(\dots,\alpha_{i}\pm1,\dots\right)\textnormal{.}
\end{align*}
one can use 
\par{\footnotesize
\begin{align*}
 0 
= \int\frac{d^{d}k\, d^{d}l_{1}\, d^{d}l_{2}}{\left(2\pi\right)^{3d}} \frac{\partial}{\partial l_{2}^{\mu}}\frac{l_{2}^{\mu}}{\left[k^{2}-1\right]^{\alpha_{1}}\left[\left(k+l_{1}\right)^{2}-1\right]^{\alpha_{2}}\left[\left(k+l_{1}+l_{2}\right)^{2}-1\right]^{\alpha_{3}}\left[l_{1}^{2}\right]^{\alpha_{4}}\left[l_{2}^{2}\right]^{\alpha_{5}}\left[\left(l_{1}+l_{2}\right)^{2}\right]^{\alpha_{6}}}\textnormal{.}
\end{align*}}%
to obtain
\begin{align*}
 I\left(\alpha_{1},\dots,\alpha_{6}\right) 
\quad= \quad  
\frac{  \left\{ \alpha_{3}\left(3^{+}\right)\left[\left(2^{-}\right)-\left(5^{-}\right)\right]+\alpha_{6}\left(6^{+}\right)\left[\left(4^{-}\right)-\left(5^{-}\right)\right]\right\}     }
{\left[\alpha_{3}+2\alpha_{5}+\alpha_{6}-D\right]}
\quad	 I\left(\alpha_{1},\dots,\alpha_{6}\right)\textnormal{.}
\end{align*}
This reduces $\alpha_{2}+\alpha_{4}+\alpha_{5}$. The relation can
be used until one of the indices is $0$. In the cases $\alpha_{4}<0$
or $\alpha_{5}<0$ this relation cannot be used. However, because of
the conventional ordering of the indices, this case would imply a
negative value of $\alpha_{6}$ and it would require $\alpha_{5}$
to be the second non-positive index. If one finally arrives at $\alpha_{4}=0$ or $\alpha_{5}=0$
, one can stop, if $\alpha_{6}$ is already non-positive. If this is
not the case, one can interchange the indices to make $\alpha_{6}$
 the index which is $0$. Then one can use the relation:
\begin{align*}
 & I\left(\alpha_{1},\alpha_{2},\alpha_{3},\alpha_{4},\alpha_{5},0\right) \quad= \quad \frac{\left\{ \alpha_{3}\left(3^{+}\right)\left[\left(2^{-}\right)-\left(5^{-}\right)\right]\right\}}{\left[\alpha_{3}+2\alpha_{5}-D\right]} I\left(\alpha_{1},\alpha_{2},\alpha_{3},\alpha_{4},\alpha_{5},0\right)\textnormal{.}
\end{align*}
This can be used until $\alpha_{2}$ or $\alpha_{5}$ is $0$. We conclude that in any case  we end up with one of the special cases given above.\\
Now we show how to compute the integrals for the special values of the coefficients.

\subparagraph{$\alpha_{1}\leq0,\alpha_{2}\leq0$ or  $\alpha_{3}\leq0$:}$\;$\\
The cases $\alpha_{1}\leq0,\alpha_{2}\leq0$ can be reduced to the case $\alpha_{3}\leq0$. In this case one has the integral
\begin{align*}
 & I\left(\alpha_{1},\alpha_{2},-\left|\alpha_{3}\right|,\alpha_{4},\alpha_{5},\alpha_{6}\right) \\
= \quad& \int\frac{d^{d}k\, d^{d}l_{1}\; d^{d}l_{2}}{\left(2\pi\right)^{2d}}\frac{\left[\left(k+l_{1}\right)^{2}+2k\cdot l_{2}+2l_{1}\cdot l_{2}+l_{2}^{2}\right]^{\left|\alpha_{3}\right|}}{\left[k^{2}-1\right]^{\alpha_{1}}\left[\left(k+l_{1}\right)^{2}-1\right]^{\alpha_{2}}\left[l_{1}^{2}\right]^{\alpha_{4}}\left[l_{2}^{2}\right]^{\alpha_{5}}
\left[\left(l_{1}+l_{2}\right)^{2}\right]^{\alpha_{6}}}\textnormal{.}
\end{align*}
If one of the indices $\alpha_{5}$ or $\alpha_{6}$
is nonnegative, the integral is 0. Expanding the numerator and doing
the obvious cancellations, one is left with integrals of the type:
\begin{align*}
 & \left(\int\frac{d^{d}k\, d^{d}l_{1}}{\left(2\pi\right)^{2d}}\frac{1}{\left[k^{2}-1\right]^{\alpha_{1}}\left[\left(k+l_{1}\right)^{2}-1\right]^{\alpha_{2}}\left[l_{1}^{2}\right]^{\alpha_{4}}}\right. 
 \left.\left(\int\frac{d^{d}l_{2}}{\left(2\pi\right)^{d}}\frac{\left(2k\cdot l_{2}\right)^{p}}{\left[l_{2}^{2}\right]^{\alpha_{5}}\left[\left(l_{1}+l_{2}\right)^{2}\right]^{\alpha_{6}}}\right)\right)\\
 = & \quad\sum_{r=0}^{\left[\frac{p}{2}\right]}C\left(\alpha_{5},\alpha_{6},r\right)\quad\int\frac{d^{d}k\, d^{d}l_{1}}{\left(2\pi\right)^{2d}}\frac{\left[l_{1}^{2}\right]^{r}\left[k^{2}\right]^{r}\left[\left(k+l_{1}\right)^{2}-k^{2}-l_{1}^{2}\right]^{p-2r}}{\left[k^{2}-1\right]^{\alpha_{1}}\left[\left(k+l_{1}\right)^{2}-1\right]^{\alpha_{2}}\left[l_{1}^{2}\right]^{\alpha_{4}+\alpha_{5}+\alpha_{6}-d/2}}
\intertext{with}
 & C\left(\alpha_{5},\alpha_{6},r\right)\quad=\quad  
\frac{i(-1)^{d/2}}{\left(4\pi\right)^{d/2}}\frac{\Gamma\left(\alpha_{5}+\alpha_{6}-\frac{d}{2}-r\right)\Gamma\left(p+\frac{d}{2}-\alpha_{6}-r\right)\Gamma\left(\frac{d}{2}-\alpha_{5}+r\right)p!}
{\Gamma\left(\alpha_{5}\right)\Gamma\left(\alpha_{6}\right)\Gamma\left(p+d-\alpha_{5}-\alpha_{6}\right)r!\left(p-2r\right)!}\textnormal{.}
\end{align*}
After the some cancellations, one has integrals of the type
\begin{align*}
 & \int\frac{d^{d}k\, d^{d}l_{1}}{\left(2\pi\right)^{2d}}\frac{1}{\left[k^{2}-1\right]^{\alpha_{1}}\left[\left(k+l_{1}\right)^{2}-1\right]^{\alpha_{2}}\left[l_{1}^{2}\right]^{\alpha_{3}}}\textnormal{.}
\end{align*}
If $\alpha_{1}$ or $\alpha_{2}$ are non-positive, the integral is
0. If all indices are positive, this is the standard-integral from
the appendix. If $\alpha_{3}$ is negative, one can actually use the
result (\ref{eq:2loop_master_integral}) as trivial analytic continuation to $\alpha_{3}<0$.

\subparagraph{$\alpha_{5}\leq0$ and $\alpha_{6}\leq0$:}$\;$\\
Renaming and exchanging integration variables, we have:
\begin{align*}
 & I\left(\alpha_{1},\alpha_{2},\alpha_{3},\alpha_{4},-\alpha_{5},-\alpha_{6}\right) \\
=&\quad \int\frac{d^{d}k\, d^{d}l_{1}\, d^{d}l_{2}}{\left(2\pi\right)^{3d}}\quad\frac{\left[l_{2}^{2}\right]^{\alpha_{5}}\left[\left(l_{1}+l_{2}\right)^{2}\right]^{\alpha_{6}}}{\left[k^{2}-1\right]^{\alpha_{1}}\left[\left(k+l_{1}\right)^{2}-1\right]^{\alpha_{2}}\left[\left(k+l_{1}+l_{2}\right)^{2}-1\right]^{\alpha_{3}}\left[l_{1}^{2}\right]^{\alpha_{4}}}\\
 =& \quad \int\frac{d^{d}k\, d^{d}l_{1}\, d^{d}l_{2}}{\left(2\pi\right)^{3d}}\quad\frac{\left[l_{1}^{2}+l_{2}^{2}+2l_{1}\cdot l_{2}\right]^{\alpha_{5}}\left[l_{2}^{2}+k^{2}+2l_{2}\cdot k\right]^{\alpha_{6}}}{\left[k^{2}-1\right]^{\alpha_{1}}\left[l_{1}^{2}-1\right]^{\alpha_{2}}\left[l_{2}^{2}-1\right]^{\alpha_{3}}\left[\left(l_{1}+k\right)\right]^{\alpha_{4}}}\textnormal{.}
\end{align*}
If any of the massive lines have non-positive indices, the integral
is 0. If $\alpha_{4}$ is non-positive, the integral reduces to a product
of three integrals like $\int\frac{d^{d}k}{\left(2\pi\right)^{d}}\frac{2l_{i}\cdot k}{\left[k^{2}-1\right]^{\alpha}}$
which can be easily done. If all indices are positive:
\begin{align*}
 & I\left(\alpha_{1},\alpha_{2},\alpha_{3},\alpha_{4}\right)\\
 =\quad& \left(\int\frac{d^{d}k\, d^{d}l_{1}}{\left(2\pi\right)^{2d}}\frac{1}{\left[k^{2}-1\right]^{\alpha_{1}}\left[l_{1}^{2}-1\right]^{\alpha_{2}}\left[\left(l_{1}+k\right)\right]^{\alpha_{3}}}\left(\int\frac{d^{d}l_{2}}{\left(2\pi\right)^{2d}}\frac{\left[2l_{1}\cdot l_{2}\right]^{p}\left[2k\cdot l_{2}\right]^{q}}{\left[l_{2}^{2}-1\right]^{\alpha_{4}}}\right)\right)\textnormal{.}
\end{align*}
Doing the $l_{2}$-integration and simplifying one has the easy integrals
(take $p<q$):
\begin{align*}
 & I\left(\alpha_{1},\alpha_{2},\alpha_{3},\alpha_{4}\right) \quad=\quad  \sum_{t=0}^{\left[\frac{p}{2}\right]}D\left(\alpha_{4},p,q,t\right)  \int\frac{d^{d}k\, d^{d}l_{1}}{\left(2\pi\right)^{2d}}\frac{\left[l_{1}^{2}\right]^{t}\left[k^{2}\right]^{\frac{q-p}{2}+t}\left[2k\cdot l_{_{1}}\right]^{p-2t}}{\left[k^{2}-1\right]^{\alpha_{1}}\left[l_{1}^{2}-1\right]^{\alpha_{2}}\left[\left(l_{1}+k\right)\right]^{\alpha_{3}}}\textnormal{.}
\intertext{Here,}
 & D\left(\alpha_{4},p,q,t\right) \quad= \quad \frac{(-1)^{\alpha_{4}+\frac{\left(p+q\right)}{2}}i}{\left(4\pi\right)^{d/2}}\quad\frac{\Gamma\left(\alpha_{4}-\frac{\left(p+q\right)}{2}-\frac{d}{2}\right)}{\Gamma\left(\alpha_{4}\right)}\quad\frac{p!\left(p+q\right)!}{t!\left(p-2t\right)!\left(\frac{q-p}{2}+t\right)!}\textnormal{.}
\end{align*}
Note that in this case all terms in the numerator can be canceled
since all exponents are integers. This gives:
\begin{align*}
 & I\left(\alpha,\beta,\gamma\right) \quad =\quad \int\frac{d^{d}k\, d^{d}l_{1}}{\left(2\pi\right)^{2d}}\frac{1}{\left[k^{2}-1\right]^{\alpha}\left[l_{1}^{2}-1\right]^{\beta}\left[\left(l_{1}+k\right)\right]^{\gamma}}\textnormal{.}
\end{align*}
This is again Eq.~ (\ref{eq:2loop_master_integral}).

\subsubsection{Results for the three-loop contributions to the vector current}
The techniques described above could be used to compute the contributions for the vector current. We use $\alpha_s=\frac{g_s}{4\pi}$ to obtain:
\begin{align*}
\left(I_{VC}\right)_{3\;loop}+\textnormal{perm.}
\quad=\quad&\left(\frac{\alpha_s}{\pi}\right)^3\frac{5  e^{i q\cdot x}}{216 M^2}\bigg\{i \epsilon^{\mu\nu \tau\rho}q_\nu p_\tau  \gamma^5\gamma_\rho\left(\frac{41}{8}-2\zeta(3)\right)\\
&-i\,m\,q_\nu \sigma^{\mu \nu}\left(\frac{49}{4}-11\,\zeta(3)\right)
-\left(q^2g^{\mu\nu}-q^{\mu}q^{\nu}\right)\gamma_\nu\left(\frac{33}{16}-\frac{10\,\zeta(3)}{3}\right)\bigg\}\textnormal{.}
\end{align*}
Here we use antisymmetrization \emph{including} the symmetry factors, for example: $q^{\left[\mu\right.}\gamma^{\left.\nu\right]}=\frac{1}{2}\left(q^\mu\gamma^\nu-q^\nu \gamma^\mu\right)$.  
In this case it is possible to read off the corresponding operators immediately.
\begin{align*}
O^{VC}_{3\;loop}\quad=\quad&
 \left(\frac{\alpha_s}{\pi}\right)^3\frac{5 }{216  M^2}\bigg\{i  \epsilon^{\mu\tau\nu \rho}\partial_{\nu}\left(\overline{\Psi}\partial_{\tau}  \gamma^5\gamma_\rho\Psi\right)\left(\frac{41}{8}-2\zeta(3)\right)\\
&-\,m\,\partial_\nu \left(\overline{\Psi}\sigma^{\mu \nu}\Psi\right)\left(\frac{49}{4}-11\,\zeta(3)\right)
+\left(\partial^2g^{\mu\nu}-\partial^{\mu}\partial^{\nu}\right)\left(\overline{\Psi}\gamma_\nu\Psi\right)\left(\frac{33}{16}-\frac{10\,\zeta(3)}{3}\right)\bigg\}\textnormal{.}
\end{align*}
Note that we can freely interchange derivatives with covariant ones since the difference is of higher order in $g_s$.
Using the QCD equations of motion, $\slashed{\nabla}\;\Psi\;=\;-i m\;\Psi$ and $\overline{\Psi}\;\<-{\slashed{\nabla}}\;=\;i m\; \overline{\Psi}$  
and the fact that the vector current for the light quarks is conserved, $\partial_\mu\;\overline{\Psi}\gamma^\mu \Psi\; =\;0$, we have 
\begin{alignat*}{4}
 -i \epsilon^{\mu\nu\tau \rho} \partial_{\nu}\left(\overline{\Psi}\partial_{\tau} \gamma^5\gamma_\rho\Psi\right)
\quad =\quad& && - m\;\partial_\nu\; \overline{\Psi}\sigma^{\mu \nu}\Psi	
                    -\frac{1}{2}\;\partial^2\left(\overline{\Psi}\gamma^\mu  \Psi\right) 
            +\mathcal{O}\left(g_s\right)\textnormal{.}
\end{alignat*}
This gives the result:
\begin{align*}
O^{VC}_{3\;loop}\quad=\quad&
\left(\frac{\alpha_s}{\pi}\right)^3 \frac{5 }{36  M^2}\bigg\{
\,m\,\partial_\nu \left(\overline{\Psi}\sigma^{\mu \nu}\Psi\right)\left(\frac{3}{2}\,\zeta(3)-\frac{19}{16}\right)
\;+\;\partial^2\left(\overline{\Psi}\gamma^\mu\Psi\right)\left(\frac{37}{48}-\frac{13\,\zeta(3)}{18}\right)\bigg\}\textnormal{.}
\end{align*}

\subsubsection{Results for the three-loop contributions to the tensor current}
Here the computation yields the result:
\begin{align*}
\left(I_{TC}\right)_{3\;loop}+\textnormal{perm.}\quad=\quad&
\left(\frac{\alpha_s}{\pi}\right)^3\frac{5 i e^{i q\cdot x}}{36 M}\bigg\{i \epsilon^{\mu\nu \tau \rho}\left(p_\tau+\frac{1}{2}q_\tau\right)\gamma^5\gamma_\rho\left(\frac{1}{3}-\zeta(3)\right)\\
&+ i\,m\, \sigma^{\mu \nu}\left(\frac{19}{6}-3 \,\zeta(3)\right)
+q^{\left[\mu\right.}\gamma^{\left.\nu\right]}\left(\frac{1}{2}+ \,\zeta(3)\right)\bigg\}\textnormal{.}
\end{align*}
In terms of operators, we have:
\begin{align*}
O^{TC}_{3\;loop}\quad=\quad&
\left(\frac{\alpha_s}{\pi}\right)^3\frac{5 }{36  M}\bigg\{\frac{i}{2} \epsilon^{\mu\nu \tau \rho}\partial_\tau \left(\overline{\Psi} \gamma^5\gamma_\rho\Psi\right)\left(\frac{1}{3}-\zeta(3)\right)
-i \epsilon^{\mu\nu \tau \rho} \overline{\Psi}\partial_\tau\gamma^5\gamma_\rho\Psi\left(\frac{1}{3}-\zeta(3)\right)\\
&- \,m\, \overline{\Psi}\sigma^{\mu \nu}\Psi\left(\frac{19}{6}-3 \,\zeta(3)\right)
+\;\partial^{\left[\mu\right.}\left(\overline{\Psi}\gamma^{\left.\nu\right]}\Psi\right)\left(\frac{1}{2}+ \,\zeta(3)\right)\bigg\}\textnormal{.}
\end{align*}
One can use the equations of motion and standard relations for the $\gamma$-matrices to show: 
\begin{align*}
-i \epsilon^{\mu\nu\tau\rho} \;\overline{\Psi}\partial_\tau \gamma^5\gamma_\rho\Psi\;&=\;
      - m \overline{\Psi}\sigma^{\mu\nu}\Psi
       +\partial^{\left[\mu\right.}\overline{\Psi}\gamma^{\left.\nu\right]}\Psi
         -\frac{i}{2}\; \epsilon^{\mu\nu\tau\rho} \partial_\tau\;\overline{\Psi} \gamma^5\gamma_\rho\Psi  +\mathcal{O}\left(g_s\right)\textnormal{.}
\end{align*}
The result for the tensor current can now be written as: 
\begin{align*}
O^{TC}_{3\;loop}\quad=\quad&
\left(\frac{\alpha_s}{\pi}\right)^3\frac{5}{36  M}\;\bigg\{- \,m\, \overline{\Psi}\sigma^{\mu \nu}\Psi\left(\frac{7}{3}-4 \,\zeta(3)\right)
+\;\frac{2}{3}\;\partial^{\left[\mu\right.}\left(\overline{\Psi}\gamma^{\left.\nu\right]}\Psi\right)\bigg\}\textnormal{.}
\end{align*}

\section{Result}
\label{sec:result}
In the final result, we have to remember that each light flavor will contribute to the three-loop diagrams. Collecting the contributions, we have
\begin{align}
\qb \gamma^\mu Q &&=  \left(\frac{\alpha_s}{\pi}\right)^3\frac{5 }{36  M^2}& \sum_{f}\bigg\{
                           m_f \partial_\nu \left(\overline{\Psi}_f\sigma^{\mu \nu}\Psi_f\right)\left(\frac{3}{2}\zeta(3)-\frac{19}{16}\right)
                           +\partial^2\left(\overline{\Psi}_f\gamma^\mu\Psi_f\right)\left(\frac{37}{48}-\frac{13\,\zeta(3)}{18}\right)\bigg\}
                           \nonumber\\
                           &&-\frac{1}{72 \pi ^2 M^4 }\;\partial_\rho  \; tr_c &\left(\frac{7}{5}\acom[F^{\mu \gamma},F_{ \gamma \delta}]
                             F^{\delta \rho }+F_{\delta \gamma}F^{\delta\gamma } F^{\mu \rho}\right)\;+\;\mathcal{O}\left(\frac{g_s^3}{M^6}\, , \,\frac{g_s^7}{M^4}\right)
                             \label{eq:mainvec}\\
\qb \sigma^{\mu\nu} Q &&=\left(\frac{\alpha_s}{\pi}\right)^3\frac{5}{36 M}
                            \sum_{f} \bigg\{&- \,m_f\, \overline{\Psi}_f\sigma^{\mu \nu}\Psi_f\left(\frac{7}{3}-4 \,\zeta(3)\right)+\;\frac{2}{3}\;\partial^{\left[\mu\right.}\left(\overline{\Psi}_f\gamma^{\left.\nu\right]}\Psi_f\right)\bigg\}
                            \nonumber\\ 
                           &&-\frac{1}{24 \pi ^2 M^3 }  tr_c& \left(\vphantom{\frac{7}{5}}F^{\gamma \nu}F^{\rho \mu}F_{ \gamma \rho}-F^{\gamma \mu}F^{\rho \nu}F_{\gamma \rho}
                             +F_{ \gamma\rho}F^{\gamma \rho}F^{\mu \nu}\right)+\mathcal{O}\left(\frac{g_s^3}{M^5}\, , \,\frac{g_s^7}{M^3}\right) \textnormal{.}
 \label{eq:maintensor}
 \end{align}

Here $\sum_{f}$ denotes the sum over all flavors of quarks that are light compared to $M$.

These operator identities can be used to analize intrinsic heavy quark content of light hadrons. One of immediate applications is the study of the intrinsic charm influence on the nucleon e.m. and tensor form factors.

\section{Application: Heavy quark contribution to magnetic moment and electromagnetic radii of the nucleon}
\label{sec:appl1}
We can use the results of the previous section to derive a relation for the gluonic contribution to the nucleons charge radius that is due to heavy quarks.
We denote the momentum transfer by $q=p^{\prime}-p$ and the sum of the momenta by $P= p^{\prime}+p$.
First of all, we need the form factor decomposition for the vector and the tensor current. The decomposition for the vector current is standard:
\begin{align}
 \bra{p',\sigma'}\overline{\Psi}\gamma^\mu \Psi(0)\ket{p,\sigma}\;=&\;\ub \left\{\gamma^\mu F_1\left(q^2\right)+i\sigma^{\mu\nu}\frac{q^\nu}{2m_N}F_2\left(q^2\right)\right\}\u  
\label{eq:vector_form factor} \end{align}
The tensor current can be decomposed as \cite{Diehl:2001pm}:
\begin{align}
 &\bra{p',\sigma'}\overline{\Psi} \;i\sigma^{\mu\nu}\Psi(0)\ket{p,\sigma}\nonumber\\
\;=&\;\ub \bigg\{i \sigma^{\mu\nu}H_{T}\left(q^2\right)+\frac{\gamma^\mu q^\nu-\gamma^\nu q^\mu}{2 m_N} E_T\left(q^2\right)
                                                 +\frac{P^{\mu}q^{\nu}-P^{\nu}q^{\mu}}{2 m_N^2}\tilde{H}_T\left(q^2\right)\bigg\}\u\textnormal{.}\label{eq:tensor_form factor} 
\end{align}
We actually need $q_\nu \bra{p',\sigma'}\overline{\Psi} \;i\sigma^{\mu\nu}\Psi(0)\ket{p,\sigma}$:
\begin{align*}
 &q_\nu \bra{p',\sigma'}\overline{\Psi} \;i\sigma^{\mu\nu}\Psi(0)\ket{p,\sigma}\\
=\;&\ub \left\{i q_\nu \sigma^{\mu\nu}H_T\left(q^2\right)+ \frac{q^2}{2m_N} \gamma^\mu E_T\left(q^2\right)+\frac{q^2}{ 2 m_N^2}P^{\mu} \tilde{H}_{T}\left(q^2\right) \right\}\u\\
=\;&\ub \left\{i q_\nu \sigma^{\mu\nu}\left(H_T\left(q^2\right)-\frac{q^2}{2 m_N^2} \tilde{H}_{T}\left(q^2\right)\right)+\frac{q^2}{2m_N}\gamma^\mu\;\kappa_{T}\left(q^2\right)\right\}\u\textnormal{.}
\end{align*}
Here we introduced the tensor magnetic form factor  $\kappa_{T}\left(q^2\right)= E_T\left(q^2\right)+2 \tilde{H}_{T}\left(q^2\right) $ and used the Gordon-identity,

Additionally we introduce form factors of the gluon operator $S^{\mu\nu}=tr_c \bigg[\frac{7}{5}\acom[F^{\mu \gamma},F_{ \gamma \delta}]F^{\delta \nu }+F_{\delta \gamma}F^{\delta\gamma } F^{\mu \nu}\bigg]$.
Since this operator is an antisymmetric tensor of rank 2,
it can be decomposed as the tensor current. Thus,
\begin{align}
&\bra{p',\sigma'}S^{\mu\nu }(0)\ket{p,\sigma}
\nonumber \\=&\; -i m_N^3\;\ub \bigg\{ i \sigma^{\mu\nu}S\left(q^2\right)+
                                                 \frac{\gamma^\mu q^\nu-\gamma^\nu q^\mu}{2 m_N} R\left(q^2\right)+\frac{P^{\mu}q^{\nu}-P^{\nu}q^{\mu}}{2 m_N^2}\tilde{S}\left(q^2\right)\bigg\}\u
                                                 \label{eq:gluonffdef1}\\ 
\intertext{and}
 & q_\nu \bra{p',\sigma'}S^{\mu\nu }(0)\ket{p,\sigma}
 \nonumber
 \\=&\;m_N^3 \;\ub \left\{q_\nu \sigma^{\mu\nu}\left(S\left(q^2\right)-\frac{q^2}{2 m_N^2} \tilde{S}\left(q^2\right)\right)
 +\frac{q^2 }{2i m_N}\gamma^\mu\left(R\left(q^2\right)+2\tilde{S}\left(q^2\right)\right)\right\}\u\textnormal{.}
 \nonumber
\end{align}

Now, we consider the nucleon matrix element of the  vector current made of heavy quarks.Using the derived operator identity (\ref{eq:mainvec}) we obtain
the heavy quark contribution to the electric form factor $G_E^Q\left(q^2\right)=F_1^Q\left(q^2\right)-\frac{q^2}{4m_N^2}F_2^Q\left(q^2\right)$ and the magnetic form factor $G_M^Q\left(q^2\right)=F_1^Q\left(q^2\right)+F_2^Q\left(q^2\right)$:

\begin{align*}
 G_E^Q\left(q^2\right)\;=\;&\left(\frac{\alpha_s}{\pi}\right)^3\frac{5 q^2 }{36 M^2} \sum_{f}\left\{-\left(\frac{37}{48}-\frac{13\,\zeta(3)}{18}\right)G^f_E\left(q^2\right)\right. \\
& \left.+\frac{m_f}{2m_N}\left(\frac{3}{2}\,\zeta(3)-\frac{19}{16}\right)  \left( \kappa^f_{T}\left(q^2\right) -H_T^f\left(q^2\right)+\frac{q^2}{2 m^2_N} \tilde{H}_T^f\left(q^2\right) \right)\right\}\\
&  -\frac{q^2 m_N^2}{144 \pi^2 M^4 }  \;\left\{\left(1+\frac{q^2}{4 m^2_N}\right)2 \tilde{S}\left(q^2\right)+R\left(q^2\right)-S\left(q^2\right)\right\}  \\
 G_M^Q\left(q^2\right)\;=\;& \left(\frac{\alpha_s}{\pi}\right)^3\frac{5 }{36 M^2}\sum_{f}\left\{-q^2\left(\frac{37}{48}-\frac{13\,\zeta(3)}{18}\right)G^f_M\left(q^2\right) \right.
\\&\left.+ \sum_{f} \frac{m_f}{2 m_N}\left(\frac{3}{2}\,\zeta(3)-\frac{19}{16}\right) \left(q^2  E_T^f\left(q^2\right) +4m_N^2 H_T^f\left(q^2\right) \right)\right\} \\
&-\frac{m_N^2}{144 \pi ^2 M^4 }  \left\{q^2 R\left(q^2\right)+4m_N^2 S\left(q^2\right)\right\} \textnormal{.}
\end{align*}
We see that if we neglect the chiral corrections of order $\sim m_f/m_N$, than the leading $\sim \left( \frac{\alpha_s}{\pi}\right)^3 \frac{1}{M^2}$ contributions can completely be expressed in terms of singlet e.m. form factors.
The subleading in heavy quark mass corrections of order $\sim 1/M^4$ are expressed in terms of form factors of the gluon operator (\ref{eq:gluonffdef1}).

This gives the following result for the heavy quark contribution to the electric and magnetic  radii, which we define  as $\left<r_{E/M}^2\right>_Q=-6\frac{dG_{E/M}^Q}{d q^2}\bigg{\vert}_{q^2=0}$:
 \begin{align}
 \left<r_{E}^2\right>_Q =\;&  \left(\frac{\alpha_s}{\pi}\right)^3\frac{5 }{6 M^2} \left\{
\left(\frac{37}{16}-\frac{13\,\zeta(3)}{6}\right)\right. \nonumber\\
&\left.-\sum_{f} \frac{m_f}{2 m_N }\left(\frac{3}{2}\,\zeta(3)-\frac{19}{16}\right)    \bigg(\kappa^f_T(0)-H_T^f(0)\bigg)\right\}
\label{eq:rE}\\
                         &  +\frac{1}{24  \pi ^2 }\frac{m_N^2}{M^4} \left(2 \tilde{S}(0)+R(0)-S(0)\vphantom{\frac{d S}{d q^2}\bigg{\vert}_{q^2=0}}\right)\nonumber\\
\intertext{and}
\left<r_{M}^2\right>_Q=\;&       \left(\frac{\alpha_s}{\pi}\right)^3\frac{5 }{6 M^2} \sum_{f}\left\{ 
 \left(\frac{37}{48}-\frac{13\,\zeta(3)}{18}\right)\mu^f_N\right.\nonumber\\
&\left.- \frac{m_f }{2 m_N }\left(\frac{3}{2}\,\zeta(3)-\frac{19}{16}\right)   \left(E_T^f(0)+4 m_N^2\frac{d H_T^f}{d q^2}\bigg{\vert}_{q^2=0}\right)\right\}
\label{eq:rM}\\
                            &+\frac{1}{24  \pi ^2 }\frac{m_N^2}{M^4} \left(R(0)+4 m_N^2\frac{d S}{d q^2}\bigg{\vert}_{q^2=0}\right)\textnormal{.} 
                            \nonumber
\end{align}

Finally, the heavy quark contribution to the magnetic moment (in nucleon magnetons) is:
\begin{align}
\label{eq:muQ}
 {\mu}_Q \;&=\;\left(\frac{\alpha_s}{\pi}\right)^3\frac{5 m_N}{18 M^2}\left(\frac{3}{2}\,\zeta(3)-\frac{19}{16}\right) \sum_{f} 
                               m_f    H_T^f\left( 0\right) 
                               -\frac{m_N^4}{36\pi ^2 M^4 }  S\left( 0\right)
\end{align}
We see that the leading $\sim 1/M^2$ contribution to $\mu_Q$ has additional suppression by the light quark mass $\sim m_f/m_N$ indicating that the subleading correction $\sim 1/M^4$
can be numerically more important than the leading one.
The result for the heavy quark contribution to the magnetic moments (\ref{eq:muQ}) coincides with the result obtained in Ref.~\cite{Ji:2006vx}, the results for the e.m. radii (\ref{eq:rE},\ref{eq:rM})
are new. 

From Eqs.~(\ref{eq:rE},\ref{eq:rM}), neglecting the corrections of order $\sim m_f/m_N$, one obtains the following model independent limits for the e.m. radii of the nucleon:
\begin{align}
\lim_{M\to\infty} M^2\left<r_{E}^2\right>_Q =\;&  \left(\frac{\alpha_s}{\pi}\right)^3\frac{5 }{12 } 
\left(\frac{37}{8}-\frac{13\,\zeta(3)}{3}\right) +O\left( \frac{m_f}{m_N}\right),
\label{eq:rElim}\\
\lim_{M\to\infty} M^2\left<r_{M}^2\right>_Q =\;&       \left(\frac{\alpha_s}{\pi}\right)^3\frac{5 }{36}  
 \left(\frac{37}{8}-\frac{13\,\zeta(3)}{3}\right)\sum_{f} \mu^f_N+O\left( \frac{m_f}{m_N}\right).
\label{eq:rMlim} 
\end{align}
We see that the limiting values of the heavy quark contribution to nucleon e.m. radii are always negative, so the intrinsic heavy quarks shrink the e.m. size of the nucleon. 
These limiting expression can be useful for lattice QCD simulations.

Now we give numerical results for the heavy quark contributions to magnetic moment and e.m. radii  in the case of the charm quark. 
We use the quark masses $m_u=2.3$~MeV, $m_d=4.8$~MeV, $m_s=95$~MeV and $m_c=1.275 $~GeV \cite{Beringer:1900zz} and the coupling constant $\alpha_s\left(m_c^2\right)=0.3$.
The tensor form factors and the tensor anomalous magnetic moments were computed in the chiral quark soliton model 
\cite{Kim:1995bq,Kim:1996vk}. The values for the proton and neutron are
$H^u_{T,p}(0)=H^d_{T,n}(0)=0.92$, $H^d_{T,p}(0)=H^u_{T,n}(0)=-0.27$, $\kappa^u_{T,p}(0)=\kappa^d_{T,n}(0)=3.03$
and  $\kappa^d_{T,p}(0)=\kappa^u_{T,n}(0)=1.56$ \cite{Ledwig:2010tu,Ledwig:2010zq}. Here isospin symmetry was assumed and the values were  evolved from a renormalization scale of $\mu^2=0.36 GeV^2$ 
to  $\mu^2=m_c^2$ by the method described in these publications.
The numerical result  for the charm quark contribution to the e.m. radii of the nucleon is:
\begin{align}
 \left<r_{E}^2\right>_{p,n}^Q =\;& \left(-5.1+54.8\;\left( R(0)-S(0)+2 \tilde{S}(0)\right)\right)\; 10^{-6} {\rm fm}^2, \\
 \left<r_{M}^2\right>_{p,n}^Q =\;& \left(-1.4+54.8\;\left( R(0)+4 m_N^2\frac{d S}{d q^2}\bigg{\vert}_{q^2=0}\right)\right)\; 10^{-6} {\rm fm}^2, 
\end{align}
Here we have the same result for protons and neutrons since the isospin violating effects due to mass difference of $u$ and $d$ quarks are negligible.
We see that if the form factors of the gluon operator (\ref{eq:gluonffdef1}) are numerically of order unity, than the subleading term $\sim 1/M^4$ is larger than the
leading term for the case of the charm quark. This stresses the importance of the estimates of the gluon form factors (\ref{eq:gluonffdef1}). One of possibilities to estimate these
form factors is the theory of instanton vacuum \cite{Diakonov:1995qy,Balla:1997hf}. However, one can easily see  that the gluon operator (\ref{eq:gluonffdef1}) is exactly zero
on the instanton field. This implies that one has to consider the contribution of the instanton--anti-instanon pair to this operator, hence the gluon form factors (\ref{eq:gluonffdef1}) 
are suppressed by the instanton packing fraction in the vacuum.  Given such suppression, we can very roughly estimate the size of the gluon form factors as:
\begin{align}
\label{eq:gluonffocenka}
{\rm gluon\ form\ factor} \sim \frac{N_c-2}{N_c}\ \frac{\pi \rho^2}{R^2}\ f_S^{(2)}\sim 10^{-2}.
\end{align} 
Here the factor $(N_c-2)$ takes into account that the gluon operator (\ref{eq:gluonffdef1}) is zero for the case of $N_c=2$, the factor $\frac{\pi \rho^2}{R^2} \sim \frac 13$ is the instanton packing fraction,
and $f^{(2)}_S\sim 0.1$ is the twist-4 contribution to the nucleon structure functions (see details in Refs.~\cite{Franz:2000ee,Polyakov:1998rb,Franz:1998hw}). 

Surely, the estimate (\ref{eq:gluonffocenka}) is very rough. Nevertheless, it indicates that suppression of the gluon forma factors by the instanton packing fraction
can be not enough to make the leading $\sim 1/M^2$ term in charm quark mass expansion
dominant numerically. Therefore, it is important to make more accurate calculations of the gluon form factors,
we shall give detailed estimate of them elsewhere. Let us note obvious general properties of the gluon form factors (\ref{eq:gluonffdef1}): 1) they exactly zero
for the case $N_c=2$, 2) they are zero for general self-dual gluon field configurations.

For the heavy quark contribution to the magnetic moment the leading $\sim 1/M^2$ term is suppressed by the light quark mass $\sim m_f/m_N$, therefore we expect that for the
case of the charm quark the subleading contribution $\sim 1/M^4$ is dominant numerically. Indeed, if one take the values of the parameters discussed above, one obtains  
the following result:
\begin{align*}
 \mu_Q \;&=\;\left(2.0\cdot10^{-4}-0.8 \;S(0)\right)10^{-3}.
\end{align*}
Clearly the contribution of the gluon form factor is dominant, therefore it is very important to make an estimate of this form factor.

\section{Application: Heavy quark mass contribution to the tensor form factors}
\label{sec:appl2}

The nucleon tensor form factors are defined by Eq.~(\ref{eq:tensor_form factor}). The heavy quark contribution to the tensor form factors can be easily read of the operator identity (\ref{eq:maintensor})
we derived. 

The corresponding equation contains the gluon operator
 $O^{\mu\nu}:=tr_c \bigg[F^{\gamma \nu}F^{\rho \mu}F_{ \gamma \rho}-F^{\gamma \mu}F^{\rho \nu}F_{\gamma \rho}
+F_{ \gamma\rho}F^{\gamma \rho}F^{\mu \nu}\bigg]$, which nucleon matrix element we parametrize as:
\begin{align}
&\bra{p',\sigma'}O^{\mu\nu }(0)\ket{p,\sigma}
\nonumber \\
\;=&\; m_N^3  \ub \bigg\{\sigma^{\mu\nu}U\left(q^2\right)+
                                                 \frac{\gamma^\mu q^\nu-\gamma^\nu q^\mu}{2 i m_N}V\left(q^2\right)
                                                +\frac{P^{\mu}q^{\nu}-P^{\nu}q^{\mu}}{2 i m_N^2}\tilde{U}\left(q^2\right)\bigg\}\u\textnormal{.}
\label{eq:gluonff2}                                                
\end{align}

The heavy quark mass contribution to the tensor form factors is less suppressed than corresponding contribution to the vector form factors - the leading contribution is of order $\sim \left(\frac{\alpha_s}{\pi}\right)^3 \frac{1}{M}$.
The result is:
{\scriptsize
\begin{align*}
H_{T}^Q\left(q^2\right)\;=&\;-\left(\frac{\alpha_s}{\pi}\right)^3\frac{5}{36 M} \left(\frac{7}{3}-4 \,\zeta(3)\right)
                            \sum_{f}  \,m_f H_{T}^{f}\left(q^2\right)-\frac{ m_N^3}{24 M^3 \pi ^2}U\left(q^2\right)\\
 E_T^Q\left(q^2\right)\;=&\;\left(\frac{\alpha_s}{\pi}
                            \right)^3\frac{5}{36 M}\left[- \left(\frac{7}{3}-4 \,\zeta(3)\right) \sum_{f}  \,m_f E_{T}^{f}\left(q^2\right)+\frac{2}{3}m_N \sum_{f} \left(F_1^f\left(q^2\right)+F_2^f\left(q^2\right)\right)\right]\\
                            &\;-\frac{ m_N^3}{24 M^3 \pi ^2}V\left(q^2\right)\\
\tilde{H}_{T}^Q\left(q^2\right)\;=&\;\left(\frac{\alpha_s}{\pi}
                            \right)^3\frac{5}{36 M} \left[-\left(\frac{7}{3}-4 \,\zeta(3)\right) \sum_{f}  \,m_f \tilde{H}_T^{f}\left(q^2\right)-\frac{1}{3}m_N \sum_{f} F_2^f\left(q^2\right)\right]-\frac{ m_N^3}{24 M^3 \pi ^2}\tilde{U}\left(q^2\right)\\
\kappa_{T}^Q\left(q^2\right)\;=&\;\;\left(\frac{\alpha_s}{\pi}
                            \right)^3\frac{5}{36 M}\left[ -\left(\frac{7}{3}-4 \,\zeta(3)\right) \sum_{f}  \,m_f \kappa_{T}^{f}\left(q^2\right)+\frac{2}{3} m_N \sum_{f}  F_1^f\left(q^2\right)\right]\\
                              &\;-\frac{ m_N^3}{24 M^3 \pi ^2}\;\left[V\left(q^2\right)+2\tilde{U}\left(q^2\right)\vphantom{\sum_{f\in\{u,d,s\}}}\right]
\end{align*}
\scriptsize}%

Neglecting the contributions suppressed by the light quark masses $\sim m_f/m_N$ we can derive the following heavy quark mass limits for the tensor form factors:

\begin{align*}
\lim_{M\to \infty} M E_T^Q\left(q^2\right)\;=&\;\left(\frac{\alpha_s}{\pi}
                            \right)^3\frac{5 m_N}{54 } \sum_{f} \left(F_1^f\left(q^2\right)+F_2^f\left(q^2\right)\right)+O\left(\frac{m_f}{m_N}\right),\\
 \lim_{M\to \infty} M   \tilde{H}_{T}^Q\left(q^2\right)\;=&\;-\left(\frac{\alpha_s}{\pi}
                            \right)^3\frac{5 m_N}{108 }  \sum_{f} F_2^f\left(q^2\right)    +O\left(\frac{m_f}{m_N}\right),\\        
 \lim_{M\to \infty} M      \kappa_{T}^Q\left(q^2\right)\;=&\;\;\left(\frac{\alpha_s}{\pi}
                            \right)^3\frac{5 m_N}{54} \sum_{f}  F_1^f\left(q^2\right) +O\left(\frac{m_f}{m_N}\right).                                
\end{align*}
All these limits are expressed completely in terms of e.m. nucleon form factors, so these are model independent results.

\section{Conclusion and outlook}
Using method of expansion by subgraph we derived the heavy quark mass expansion of the vector and tensor currents up three loops. The corresponding operator identities are given
by Eqs. (\ref{eq:mainvec},\ref{eq:maintensor}). We applied these result to analyse the influence of the intrinsic charm on the nucleon e.m. and tensor form factors. We show that the leading orders of $1/M$ expansion
are model independent and can be expressed in terms of known light flavour nucleon form factors. For the subleading terms one needs the calculation of the nucleon matrix elements of the gluon operators
of dimension six. We showed that the corresponding operators in the instanton vacuum are suppressed by the instanton packing fraction. However, still the contribution of the gluon operators can be important numerically
for the case of charm quark.

The heavy quark mass expansion Eqs. (\ref{eq:mainvec},\ref{eq:maintensor}) can used to study the intrinsic heavy quark content of other hadrons, for example of vector mesons. The detailed studies of these topics
will be published elsewhere. Also the methods developed here can be easily generalized to the case of other heavy quark made fermionic operators. For example, one can derive the heavy quark mass expansion of the
twist-2 heavy quark quark operators. This gives us direct access to the intrinsic heavy quark content in the nucleon structure functions.

 \section{Acknowledgements}
 M.V.P. acknowledges support of RSF grant 14-22-0281.
 \newpage

\appendix

\section{Two loop contributions at $\mathcal{O}\left(g_s^5\right)$}\label{appendix_2loop_I}
This appendix gives the results for the two loop integrals at $\mathcal{O}\left(g_s^5\right)$ (section \ref{main:2loop}).

\begin{figure}[h]
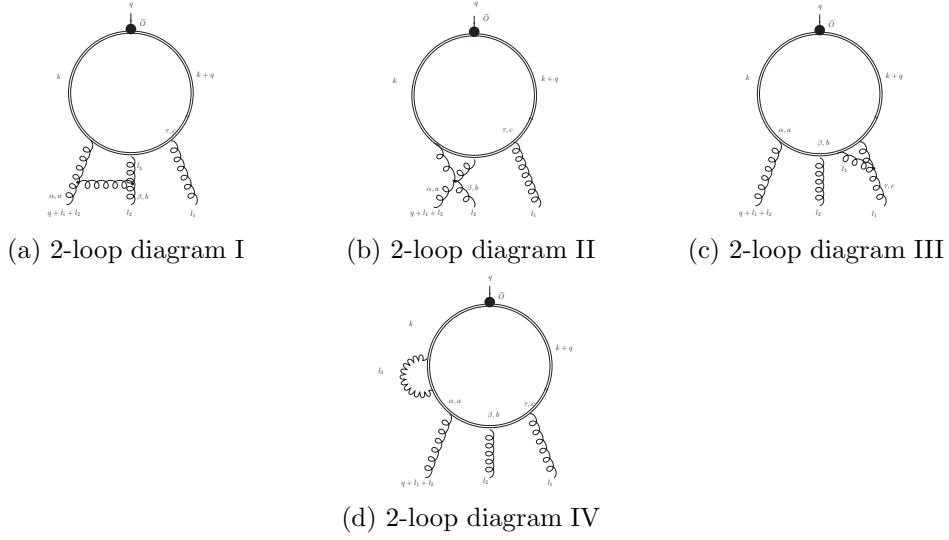

  \centering
\subcaptionbox{2-loop diagram I\label{2loop3gluon_appendixg5}}
  [.3\linewidth]{\includegraphics[scale=.2]{2loop_new.eps}}
\subcaptionbox{2-loop diagram II\label{2loop_appendixg5}}
  [.3\linewidth]{\includegraphics[scale=.2]{2loop_gluon.eps}}
\subcaptionbox{2-loop diagram III\label{2loop4gluon_appendixg5}}
  [.3\linewidth]{\includegraphics[scale=.2]{loop2_4gluon_1.eps}}
\subcaptionbox{2-loop diagram IV\label{2loop5gluon_appendixg5}}
  [.3\linewidth]{\includegraphics[scale=.2]{loop2_5gluon_1.eps}}
\caption{Two loop contributions}
\label{fig1_appendixg5}
\end{figure}
All diagrams were computed with a fixed assignment of external momenta. After the results were obtained, symmetrization in external momenta was done. Note that before symmetrization, 
the diagrams of Fig.~\ref{2loop_appendixg5} come with an additional factor of $1/2$ due to the four-gluon-vertex. To simplify the expressions, we denote the three- and four-gluon-vertices as  
\begin{align*}
&V^{\mu \nu \tau}_{abc}\left(l_1,l_2,l_3\right)\quad&=\quad &g\;f^{abc} \bigg\{g^{\mu\nu}\left(l_1-l_2\right)^\tau\;+\;g^{\nu\tau}\left(l_2-l_3\right)^\mu\;
                                                                           +\;g^{\tau\mu}\left(l_3-l_1\right)^\nu\bigg\}\\
&W^{\mu \nu \tau\rho}_{abcd}\quad&=\quad& -ig^2\bigg\{\phantom{+}f^{abe}f^{cde} \left(g^{\mu\tau}g^{\nu\rho}-g^{\mu\rho}g^{\nu\tau}\right)
                                                   +f^{ace}f^{bde} \left(g^{\mu\nu}g^{\tau\rho}-g^{\mu\rho}g^{\nu\tau}\right)\\
                                                   &&&\phantom{-ig^2\bigg\{}+f^{ade}f^{bce} \left(g^{\mu\nu}g^{\tau\rho}-g^{\mu\tau}g^{\nu\rho}\right)\bigg\}
\textnormal{.}\end{align*}
Then we have
\begin{align*}
 I^{I}_{2\;loop}\quad=\quad& -i\left(i g_s\right)^3 \int \frac{d^d k\;d^d r}{\left( 2 \pi\right)^{2d}}\;V_{\overline{a} a e}^{\overline{\alpha} \alpha \rho}\left(q+r+l_1,-q-l_1-l_2,l_2-r\right)
\;V_{\overline{b} e b}^{\overline{\beta} \rho \beta}\left(-r,r-l_2,l_2\right)\\
  &tr_c\left(T_{\overline{a}}T_{\overline{b}}T_{c}\right)\quad\frac{1}{r^2}\;\frac{1}{\left(q+r+l_1\right)^2}\;\frac{1}{\left(r-l_2\right)^2}\\
&tr_\gamma \bigg(\Gamma\; \frac{1}{\slashed{k}-M} \;\gamma^{\overline{\alpha}}\;\frac{1}{\slashed{k}+\slashed{q}+\slashed{l_1}+\slashed{r}-M}
\;\gamma^{\overline{\beta}}\;\frac{1}{\slashed{k}+\slashed{q}+\slashed{l_1}-M}
\;\gamma^{\tau}\;\frac{1}{\slashed{k}+\slashed{q}-M}\bigg)\\
I^{II}_{2\;loop}\quad=\quad& \frac{1}{2}\left(i g_s\right)^3 \int \frac{d^d k\;d^d r}{\left( 2 \pi\right)^{2d}}\;W_{\overline{a}\overline{b} b a}^{\overline{\alpha}\overline{\beta} \beta \alpha }
  tr_c\left(T_{\overline{a}}T_{\overline{b}}T_{c}\right)\quad\frac{1}{r^2}\;\frac{1}{\left(q+r+l_1\right)^2}\\
&tr_\gamma \bigg(\Gamma\; \frac{1}{\slashed{k}-M} \;\gamma^{\overline{\alpha}}\;\frac{1}{\slashed{k}+\slashed{q}+\slashed{l_1}+\slashed{r}-M}
\;\gamma^{\overline{\beta}}\;\frac{1}{\slashed{k}+\slashed{q}+\slashed{l_1}-M}
\;\gamma^{\tau}\;\frac{1}{\slashed{k}+\slashed{q}-M}\bigg)\\
I^{III}_{2\;loop}\quad=\quad& i \left(i g_s\right)^4 \int \frac{d^d k\;d^d r}{\left( 2 \pi\right)^{2d}}\;\;V_{\overline{c} e c}^{\overline{\tau} \rho\tau}\left(-r-l_1,r,l_1\right)
  tr_c\left(T_{a}T_{b}T_e T_{\overline{c}}\right)\quad\frac{1}{r^2}\;\frac{1}{\left(r+l_1\right)^2}\\
&tr_\gamma \bigg(\Gamma\; \frac{1}{\slashed{k}-M} \;\gamma^{\alpha}\;\frac{1}{\slashed{k}+\slashed{q}+\slashed{l_1}+\slashed{l_2}-M}
\;\gamma^{\beta}\;\frac{1}{\slashed{k}+\slashed{q}+\slashed{l_1}-M}\\
&\;\gamma^{\rho}\;\;\frac{1}{\slashed{k}+\slashed{q}+\slashed{l_1}+\slashed{r}-M}\;\gamma^{\overline{\tau}}\;\frac{1}{\slashed{k}+\slashed{q}-M}\bigg)\\
I^{IV}_{2\;loop}\quad=\quad&- i \left(i g_s\right)^5 \int \frac{d^d k\;d^d r}{\left( 2 \pi\right)^{2d}}\;tr_c\left(T_{e}T_{e}T_{a}T_{b}T_{c}\right)\quad\frac{1}{r^2}\\
&tr_\gamma \bigg(\Gamma\; \frac{1}{\slashed{k}-M} \;\gamma_{\rho}\; \frac{1}{\slashed{k}+\slashed{r}-M} \;\gamma^{\rho}\;\frac{1}{\slashed{k}-M} \;\gamma^{\alpha}
\;\frac{1}{\slashed{k}+\slashed{q}+\slashed{l_1}+\slashed{l_2}-M}\;\\
&\gamma^{\beta}\;\frac{1}{\slashed{k}+\slashed{q}+\slashed{l_1}-M}\;\gamma^{\tau}\;\frac{1}{\slashed{k}+\slashed{q}-M}\bigg)
\textnormal{.}\end{align*}
We have three different integrals of type I, three different integrals of type II, six integrals of type III and ten different integrals of type IV. However, all
integrals in one class, can in spite of there different topologies, be computed with the same methods.\par
We computed all diagrams. For both the vector and the tensor current the sum of all  contributions is $0$. This is what was expected for reasons of gauge invariance.\newpage

\section{Two loop contributions at $\mathcal{O}\left(g_s^6\right)$}\label{appendix_2loop_II}
This appendix gives the results for the two loop integrals at $\mathcal{O}\left(g_s^6\right)$ (section \ref{main:2loop}).
The computation here was very similar to that for the order $g_s^5$. Thus, here only the different types of diagrams are given, not each diagram that was computed.

\begin{figure}[h]
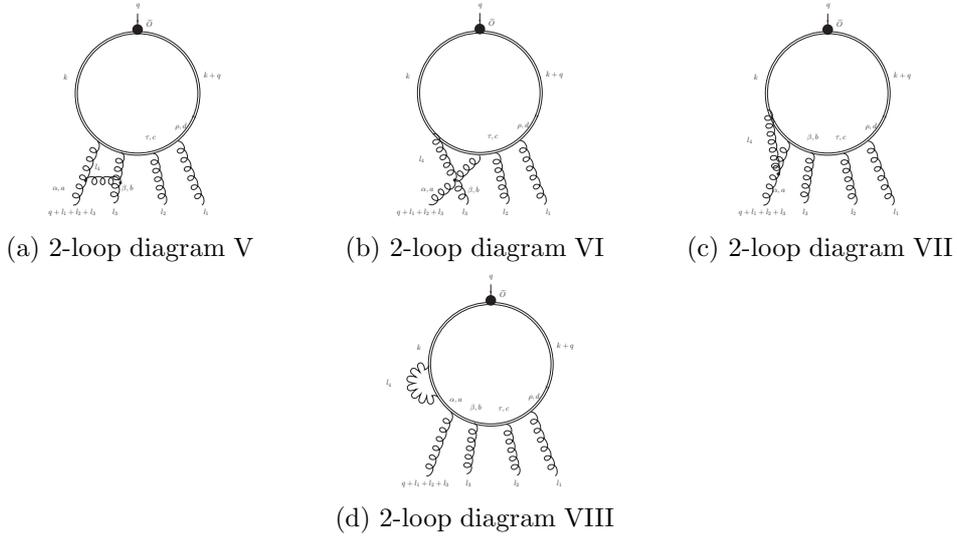

  \centering
\subcaptionbox{2-loop diagram V\label{2loop_higherorder_appendixg6}}
  [.3\linewidth]{\includegraphics[scale=.2]{2loop_4_ext_gluons.eps}}
\subcaptionbox{2-loop diagram VI\label{2loop3gluon_higherorder_appendixg6}}
  [.3\linewidth]{\includegraphics[scale=.2]{loop2_4gluon_g6.eps}}
\subcaptionbox{2-loop diagram VII\label{2loop4gluon_higherorder_appendixg6}}
  [.3\linewidth]{\includegraphics[scale=.2]{loop2_5gluon_g6.eps}}
\subcaptionbox{2-loop diagram VIII\label{2loop5gluon_higherorder_appendixg6}}
  [.3\linewidth]{\includegraphics[scale=.2]{2loop_4_ext_gluons2.eps}}
\caption{Two loop contributions}
\label{fig1_appendixg6}
\end{figure}
Again, the integrals were first done with a fixed assignment of external momenta. There are 6 diagrams of type shown in Fig.~\ref{2loop_higherorder_appendixg6},
10 diagrams of type \ref{2loop3gluon_higherorder_appendixg6}, 10 diagrams of type \ref{2loop4gluon_higherorder_appendixg6} and 15 diagrams of 
type \ref{2loop5gluon_higherorder_appendixg6}.
Finally, there are 24 permutations of external momenta. The explicit expressions for the diagrams are

\begin{align*}
 I^{V}_{2\;loop}\quad=\quad& \left(i g_s\right)^4 \int \frac{d^d k\;d^d r}{\left( 2 \pi\right)^{2d}}\;V_{\overline{a} a e}^{\overline{\alpha} \alpha \rho}\left(q+r+l_2,-q-l_2-l_3,l_3-r\right)
\;V_{\overline{b} e b}^{\overline{\beta} \rho \beta}\left(-r,r-l_3,l_3\right)\\
  &tr_c\left(T_{\overline{a}}T_{\overline{b}}T_{c}T_d\right)\quad\frac{1}{r^2}\;\frac{1}{\left(q+r+l_1\right)^2}\;\frac{1}{\left(r-l_2\right)^2}\\
&tr_\gamma \bigg(\Gamma\; \frac{1}{\slashed{k}-M} \;\gamma^{\overline{\alpha}}\;\frac{1}{\slashed{k}+\slashed{q}+\slashed{l_1}+\slashed{l_2}+\slashed{r}-M}
\;\gamma^{\overline{\beta}}\;\frac{1}{\slashed{k}+\slashed{q}+\slashed{l_1}+\slashed{l_2}-M}\\
&\gamma^{\tau}\;\frac{1}{\slashed{k}+\slashed{q}+\slashed{l_1}-M}\;\gamma^{\rho}\;\frac{1}{\slashed{k}+\slashed{q}-M}\bigg)\\
I^{VI}_{2\;loop}\quad=\quad& \frac{i}{2}\left(i g_s\right)^4 \int \frac{d^d k\;d^d r}{\left( 2 \pi\right)^{2d}}\;W_{\overline{a}\overline{b} b a}^{\overline{\alpha}\overline{\beta} \beta \alpha }
  tr_c\left(T_{\overline{a}}T_{\overline{b}}T_{c}T_{d}\right)\quad\frac{1}{r^2}\;\frac{1}{\left(q+r+l_1\right)^2}\\
&tr_\gamma \bigg(\Gamma\; \frac{1}{\slashed{k}-M} \;\gamma^{\overline{\alpha}}\;\frac{1}{\slashed{k}+\slashed{q}+\slashed{l_1}+\slashed{l_2}+\slashed{r}-M}
\;\gamma^{\overline{\beta}}\;\frac{1}{\slashed{k}+\slashed{q}+\slashed{l_1}+\slashed{l_2}-M}\\
&\gamma^{\tau}\;\frac{1}{\slashed{k}+\slashed{q}+\slashed{l_1}-M}\;\gamma^{\rho}\;\frac{1}{\slashed{k}+\slashed{q}-M}\bigg)\\
I^{III}_{2\;loop}\quad=\quad& - \left(i g_s\right)^5 \int \frac{d^d k\;d^d r}{\left( 2 \pi\right)^{2d}}\;\;V_{\overline{d} e d}^{\overline{\rho} \delta\rho}\left(-r-l_1,r,l_1\right)
  tr_c\left(T_{a}T_{b}T_c T_{e}T_{\overline{d}}\right)\quad\frac{1}{r^2}\;\frac{1}{\left(r+l_1\right)^2}\\
&tr_\gamma \bigg(\Gamma\; \frac{1}{\slashed{k}-M} \;\gamma^{\alpha}\;\frac{1}{\slashed{k}+\slashed{q}+\slashed{l_1}+\slashed{l_2}+\slashed{l_3}-M}
\;\gamma^{\beta}\;\frac{1}{\slashed{k}+\slashed{q}+\slashed{l_1}+\slashed{l_2}-M}\\
&\gamma^{\tau}\;\frac{1}{\slashed{k}+\slashed{q}+\slashed{l_1}-M}\;\gamma^{\delta}\;\;\frac{1}{\slashed{k}+\slashed{q}+\slashed{l_1}+\slashed{r}-M}
\;\gamma^{\overline{\rho}}\;\frac{1}{\slashed{k}+\slashed{q}-M}\bigg)\\
I^{IV}_{2\;loop}\quad=\quad& \left(i g_s\right)^6 \int \frac{d^d k\;d^d r}{\left( 2 \pi\right)^{2d}}\;tr_c\left(T_{e}T_{e}T_{a}T_{b}T_{c}T_{d}\right)\quad\frac{1}{r^2}\\
&tr_\gamma \bigg(\Gamma\; \frac{1}{\slashed{k}-M} \;\gamma_{\delta}\; \frac{1}{\slashed{k}+\slashed{r}-M} \;\gamma^{\delta}\;\frac{1}{\slashed{k}-M} \;\gamma^{\alpha}
\;\frac{1}{\slashed{k}+\slashed{q}+\slashed{l_1}+\slashed{l_2}+\slashed{l_3}-M}\;\\
&\gamma^{\beta}\;\frac{1}{\slashed{k}+\slashed{q}+\slashed{l_1}+\slashed{l_2}-M}\;\gamma^{\tau}\;\frac{1}{\slashed{k}+\slashed{l_1}+\slashed{q}-M}
\;\gamma^{\rho}\;\frac{1}{\slashed{k}+\slashed{q}-M}\bigg)
\textnormal{.}\end{align*}
Again, as expected from gauge invariance, for both the vector and the tensor current the sum of all four contributions is $0$.

\section{Integrals necessary for the calculation of the three-loop integrals}
Here we give the results for the master integrals needed in Section \ref{main:3loop}.
These integrals can be found in \cite{Smirnov:2004ym} and \cite{Grozin:2005yg}. 
The notation $\left\{ \left[g\right]^{i}\left[l_1\right]^{n_1}\dots\left[l_k\right]^{n_k}\right\} ^{\mu_{1}\dots\mu_{n}}$ with $2 i+\sum_{j=1}^k n_j=n$ means the sum of all possible
distributions of indices $\mu_1,\dots,\mu_n$ on $i$ metric tensors and momenta $l_i$.
\begin{align*}
&I\left(\alpha,\mu_{1}\dots\mu_{n}\right)
 \quad &=&\quad \int\frac{d^{d}k}{\left(2\pi\right)^{d}}\frac{k^{\mu_{1}}\dots k^{\mu_{n}}}{\left[k^{2}-1\right]^{\alpha}}\\
&\quad &=&\quad  \frac{(-1)^{\alpha+n/2}i}{\left(4\pi\right)^{d/2}2^{n/2}}\quad\frac{\Gamma\left(\alpha-\frac{n}{2}-\frac{d}{2}\right)}{\Gamma\left(\alpha\right)}\left\{ \left[g\right]^{n/2}\right\} ^{\mu_{1}\dots\mu_{n}}\\
&I\left(\alpha,\beta,p\right)
 \quad& =&\quad  \int\frac{d^{d}k}{\left(2\pi\right)^{d}}\frac{\left(2k\cdot q\right)^{n}}{\left[k^{2}\right]^{\alpha}\left[\left(k+p\right)^{2}\right]^{\beta}}\\
 &\quad &=& \quad \frac{i(-1)^{d/2+n}}{\left(4\pi\right)^{d/2}}\frac{\Gamma\left(n+1\right)}{\Gamma\left(\alpha\right)\Gamma\left(\beta\right)\Gamma\left(d+n-\alpha-\beta\right)}\\
 &&&\quad\sum_{t=0}^{\left[\frac{n}{2}\right]}
\frac{\Gamma\left(\alpha+\beta-t-\frac{d}{2}\right)\Gamma\left(\frac{d}{2}+n-\alpha-t\right)\Gamma\left(\frac{d}{2}+t-\beta\right)}{\Gamma\left(t+1\right)\Gamma\left(n-2t+1\right)}\frac{\left[q^{2}\right]^{t}\left[2p\cdot q\right]^{n-2t}}{\left[p^{2}\right]^{\alpha+\beta-t-d/2}}\\
& I\left(\alpha,\beta,\gamma\right) \quad &=&\quad  \int\frac{d^{d}k\, d^{d}l}{\left(2\pi\right)^{2d}}\frac{1}{\left[k^{2}-1\right]^{\alpha}\left[\left(k+l\right)^{2}-1\right]^{\beta}\left[l^{2}\right]^{\gamma}}\\
&\quad &=&\quad \frac{-(-1)^{\alpha+\beta+\gamma}}{\left(4\pi\right)^{d}}\left(\frac{\Gamma\left(\alpha+\beta+\gamma-d\right)
\Gamma\left(\alpha+\gamma-\frac{d}{2}\right)\Gamma\left(\beta+\gamma-\frac{d}{2}\right)\Gamma\left(\frac{d}{2}-\gamma\right)}{\Gamma\left(\alpha+\beta+2\gamma-d\right)\Gamma\left(\alpha\right)\Gamma\left(\beta\right)\Gamma\left(\frac{d}{2}\right)}\right)\end{align*}

\bibliographystyle{unsrt}
\bibliography{bibliography}

\begin{thebibliography}{10}

\bibitem{Brodsky:1980pb}
S.J. Brodsky, P.~Hoyer, C.~Peterson, and N.~Sakai.
\newblock {The Intrinsic Charm of the Proton}.
\newblock {\em Phys.Lett.}, B93:451--455, 1980.

\bibitem{Harris:1995jx}
B.W. Harris, J.~Smith, and R.~Vogt.
\newblock {Reanalysis of the EMC charm production data with extrinsic and
  intrinsic charm at NLO}.
\newblock {\em Nucl.Phys.}, B461:181--196, 1996.

\bibitem{Brodsky:2015fna}
S.J. Brodsky, A.~Kusina, F.~Lyonnet, I.~Schienbein, H.~Spiesberger, et~al.
\newblock {A review of the intrinsic heavy quark content of the nucleon}.
\newblock 2015.

\bibitem{Franz:2000ee}
M.~Franz, Maxim~V. Polyakov, and K.~Goeke.
\newblock {Heavy quark mass expansion and intrinsic charm in light hadrons}.
\newblock {\em Phys.Rev.}, D62:074024, 2000.

\bibitem{vc2009}
Jan Sieverding.
\newblock {Nichtperturbative Beitraege zum Vektor-Formfaktor}.
\newblock diploma thesis, Ruhr Universitaet Bochum, 2009.

\bibitem{Ji:2006vx}
Xiang-dong Ji and D.~Toublan.
\newblock {Heavy-quark contribution to the proton's magnetic moment}.
\newblock {\em Phys.Lett.}, B647:361--365, 2007.

\bibitem{Kaplan:1988ku}
David~B. Kaplan and Aneesh Manohar.
\newblock {Strange Matrix Elements in the Proton from Neutral Current
  Experiments}.
\newblock {\em Nucl.Phys.}, B310:527, 1988.

\bibitem{Laporta:1991zw}
S.~Laporta and E.~Remiddi.
\newblock {The Analytic value of the light-light vertex graph contributions to
  the electron (g-2) in QED}.
\newblock {\em Phys.Lett.}, B265:182--184, 1991.

\bibitem{Appelquist:1974tg}
Thomas Appelquist and J.~Carazzone.
\newblock {Infrared Singularities and Massive Fields}.
\newblock {\em Phys.Rev.}, D11:2856, 1975.

\bibitem{Kazama:1979xc}
Yoichi Kazama and York-Peng Yao.
\newblock {A SYSTEMATIC INVESTIGATION OF EFFECTS OF HEAVY PARTICLES VIA
  FACTORIZED LOCAL OPERATORS AND RENORMALIZATION. GROUP 1: GENERAL FORMULATION
  IN QUANTUM ELECTRODYNAMICS}.
\newblock {\em Phys.Rev.}, D21:1116, 1980.

\bibitem{Witten:1975bh}
Edward Witten.
\newblock {Heavy Quark Contributions to Deep Inelastic Scattering}.
\newblock {\em Nucl.Phys.}, B104:445--476, 1976.

\bibitem{Bernreuther:1981sg}
Werner Bernreuther and Werner Wetzel.
\newblock {Decoupling of Heavy Quarks in the Minimal Subtraction Scheme}.
\newblock {\em Nucl.Phys.}, B197:228, 1982.

\bibitem{Smirnov:1990rz}
Vladimir~A. Smirnov.
\newblock {Asymptotic expansions in limits of large momenta and masses}.
\newblock {\em Commun.Math.Phys.}, 134:109--137, 1990.

\bibitem{Smirnov:2002pj}
Vladimir~A. Smirnov.
\newblock {Applied asymptotic expansions in momenta and masses}.
\newblock {\em Springer Tracts Mod.Phys.}, 177:1--262, 2002.

\bibitem{Caswell:1981xt}
William~E. Caswell and A.D. Kennedy.
\newblock {THE ASYMPTOTIC BEHAVIOR OF FEYNMAN INTEGRALS}.
\newblock {\em Phys.Rev.}, D28:3073, 1983.

\bibitem{Mertig:1990an}
R.~Mertig, M.~Bohm, and Ansgar Denner.
\newblock {FEYN CALC: Computer algebraic calculation of Feynman amplitudes}.
\newblock {\em Comput.Phys.Commun.}, 64:345--359, 1991.

\bibitem{Anikin:1978tj}
S.A. Anikin and O.I. Zavyalov.
\newblock {Short Distance and Light Cone Expansions for Products of Currents}.
\newblock {\em Annals Phys.}, 116:135--166, 1978.

\bibitem{Gorishnii:1986gn}
S.G. Gorishnii and S.A. Larin.
\newblock {Coefficient Functions of Asymptotic Operator Expansions in Minimal
  Subtraction Scheme}.
\newblock {\em Nucl.Phys.}, B283:452, 1987.

\bibitem{Broadhurst:1991fi}
David~J. Broadhurst.
\newblock {Three loop on-shell charge renormalization without integration:
  Lambda-MS (QED) to four loops}.
\newblock {\em Z.Phys.}, C54:599--606, 1992.

\bibitem{Steinhauser:2000ry}
Matthias Steinhauser.
\newblock {MATAD: A Program package for the computation of MAssive TADpoles}.
\newblock {\em Comput.Phys.Commun.}, 134:335--364, 2001.

\bibitem{Smirnov:2008iw}
A.V. Smirnov.
\newblock {Algorithm FIRE -- Feynman Integral REduction}.
\newblock {\em JHEP}, 0810:107, 2008.

\bibitem{Diehl:2001pm}
M.~Diehl.
\newblock {Generalized parton distributions with helicity flip}.
\newblock {\em Eur.Phys.J.}, C19:485--492, 2001.

\bibitem{Beringer:1900zz}
J.~Beringer et~al.
\newblock {Review of Particle Physics (RPP)}.
\newblock {\em Phys.Rev.}, D86:010001, 2012.

\bibitem{Kim:1995bq}
Hyun-Chul Kim, Maxim~V. Polyakov, and Klaus Goeke.
\newblock {Nucleon tensor charges in the SU(2) chiral quark - soliton model}.
\newblock {\em Phys.Rev.}, D53:4715--4718, 1996.

\bibitem{Kim:1996vk}
Hyun-Chul Kim, Maxim~V. Polyakov, and Klaus Goeke.
\newblock {Tensor charges of the nucleon in the SU(3) chiral quark soliton
  model}.
\newblock {\em Phys.Lett.}, B387:577--581, 1996.

\bibitem{Ledwig:2010tu}
Tim Ledwig, Antonio Silva, and Hyun-Chul Kim.
\newblock {Tensor charges and form factors of SU(3) baryons in the
  self-consistent SU(3) chiral quark-soliton model}.
\newblock {\em Phys.Rev.}, D82:034022, 2010.

\bibitem{Ledwig:2010zq}
Tim Ledwig, Antonio Silva, and Hyun-Chul Kim.
\newblock {Anomalous tensor magnetic moments and form factors of the proton in
  the self-consistent chiral quark-soliton model}.
\newblock {\em Phys.Rev.}, D82:054014, 2010.

\bibitem{Diakonov:1995qy}
Dmitri Diakonov, Maxim~V. Polyakov, and C.~Weiss.
\newblock {Hadronic matrix elements of gluon operators in the instanton
  vacuum}.
\newblock {\em Nucl.Phys.}, B461:539--580, 1996.

\bibitem{Balla:1997hf}
J.~Balla, Maxim~V. Polyakov, and C.~Weiss.
\newblock {Nucleon matrix elements of higher twist operators from the instanton
  vacuum}.
\newblock {\em Nucl.Phys.}, B510:327--364, 1998.

\bibitem{Polyakov:1998rb}
Maxim~V. Polyakov, A.~Schafer, and O.V. Teryaev.
\newblock {The Intrinsic charm contribution to the proton spin}.
\newblock {\em Phys.Rev.}, D60:051502, 1999.

\bibitem{Franz:1998hw}
M.~Franz, P.V. Pobylitsa, Maxim~V. Polyakov, and K.~Goeke.
\newblock {On the heavy quark mass expansion for the operator anti-Q (gamma(5))
  Q and the charm content of eta, eta-prime}.
\newblock {\em Phys.Lett.}, B454:335--338, 1999.

\bibitem{Smirnov:2004ym}
Vladimir~A. Smirnov.
\newblock {Evaluating Feynman integrals}.
\newblock {\em Springer Tracts Mod.Phys.}, 211:1--244, 2004.

\bibitem{Grozin:2005yg}
A~Grozin.
\newblock {Lectures on QED and QCD}.
\newblock Technical Report hep-ph/0508242. TTP-2005-15, Karlsruhe Fridericiana
  Univ., Karlsruhe, Aug 2005.

\end{thebibliography}

\end{document}